\documentclass[%
 reprint,
 superscriptaddress,
 amsmath,
 amssymb,
 aps,
 prd,
 floatfix,
]{revtex4-2}

\usepackage{graphicx}% Include figure files
\usepackage{dcolumn}% Align table columns on decimal point
\usepackage{bm}% bold math
\usepackage{hyperref}% add hypertext capabilities
\usepackage[mathlines]{lineno}% Enable numbering of text and display math
\usepackage{xspace}
\newcommand{\burst}{\texttt{\sc{burst}}\xspace}
\newcommand{\He}{\ensuremath{^4{\mathrm{He}}}\xspace}
\newcommand{\he}{\ensuremath{^3{\mathrm{He}}}\xspace}
\newcommand{\Deu}{\ensuremath{\mathrm{D}}\xspace}
\newcommand{\Li}{\ensuremath{^7{\mathrm{Li}}}\xspace}
\newcommand{\yp}{\ensuremath{Y_{\mathrm{P}}}\xspace}
\newcommand{\np}{\ensuremath{n/p}\xspace}
\newcommand{\xd}{\ensuremath{X_{\Deu}}\xspace}
\newcommand{\xhe}{\ensuremath{X_{\he}}\xspace}
\newcommand{\xli}{\ensuremath{X_{\Li}}\xspace}
\newcommand{\tfo}{\ensuremath{T_{\mathrm{fo}}}\xspace}

\begin{document}

\preprint{arXiv:2511.18646}

\title{How Primordial Black Holes Change BBN}% Force line breaks with \\

\author{Tianning Wang}
\email{tnwang@unc.edu}
\affiliation{Department of Physics and Astronomy, UNC-Chapel Hill, NC, USA.}

\author{Evan Grohs}
\email{ebgrohs@ncsu.edu}
\affiliation{Department of Physics, North Carolina State University, NC, USA.}

\author{Laura Mersini-Houghton}
\email{mersini@physics.unc.edu}
\affiliation{Department of Physics and Astronomy, UNC-Chapel Hill, NC, USA.}

%\affiliation{%


%\affiliation{
%\affiliation{
%\affiliation{%


\date{\today}% It is always \today, today,

\begin{abstract}
Primordial Black Holes (PBHs) provide a powerful probe of the early universe physics, linking inflationary fluctuations to observable cosmological phenomena. In this work, we use a bottom-up approach to study how PBHs with masses in the range $10^{8} \leq M \leq 10^{13}\,\mathrm{g}$ modify Big Bang Nucleosynthesis (BBN) through Hawking radiation. We incorporate PBH evaporation into the \burst reaction network code to evaluate its impact on light-element abundances. Our analysis illustrates that PBH evaporation acts as an entropy injection mechanism, increasing the comoving entropy density. To reproduce the observed comoving entropy density per baryon $(s/n_{\mathrm{b}})$ from the CMB, BBN simulations must therefore begin with a smaller initial entropy than in the standard scenario without PBHs. The results also reveal a threshold near $M \approx 10^{10}\,\mathrm{g}$ that separates two distinct regimes of BBN behavior. As an example, for $M\geq10^{10}\,\mathrm{g}$, the final \He mass fraction \yp obtained from calculation increases monotonically with the initial PBH mass fraction $\beta_{30}$ specified at the starting temperature $T=30\,\mathrm{MeV}$. This trend is driven by the enhanced Hubble expansion caused by the PBH energy density. In contrast, for $M \leq 10^{10}\,\mathrm{g}$, \yp exhibits non-monotonic behavior shaped by the timing of PBH evaporation and its influence on nuclear reaction rates. These findings highlight the sensitivity of BBN to PBH evaporation and establish a framework for understanding how PBH populations influence the thermal history of the early universe.

%\begin{description}
%\end{description}
\end{abstract}

\maketitle

%\tableofcontents

%------------------------------------------------Introduction------------------------------------------------
\section{\label{sec:intro}Introduction}
There has been a growing interest in the study of Primordial Black Holes (PBHs) within the fields of astrophysics and cosmology. Unlike astrophysical black holes, whose formation is associated with stellar evolution and subsequent growth, PBHs originate from primordial density fluctuations and can therefore form well before the first stars and galaxies. The diversity in their possible masses motivates the investigation of alternative formation mechanisms for lower-mass black holes. Formed in the early universe, PBHs may arise through various processes generating primordial inhomogeneities~\cite{Carr:2020gox, Green:1997sz, Byrnes:2012yx, Kodama:1982sf, Kuhnel:2015vtw, Deng:2017uwc, Deng:2016vzb, Harada:2016mhb, Yoo:2018kvb, Harada:2017fjm, Jedamzik:1996mr, Jedamzik:1999am, Polnarev:1988dh, Caldwell:1995fu, Cotner:2019ykd, Cotner:2016cvr}.

Although the existence of PBHs remains speculative, their wide mass spectrum and potential abundances make them compelling candidates for dark matter and, as we show here, for altering the cosmic evolution. Their possible effects across different cosmological epochs therefore make them valuable probes of particle physics and the thermal history of the early universe.~\cite{Carr:2016drx, Green:2020jor, Carr:2020xqk, Bird:2016dcv, Inomata:2017okj, Clesse:2016vqa, Chisholm:2005vm, Fuller:2017uyd, Wu:2025ovd, Boccia:2024nly}.

Another motivation for studying PBHs is their potential role in the formation of cosmic structures, including galaxies and clusters. Their gravitational influence can attract surrounding matter, facilitating the hierarchical growth of large-scale structures. Theoretical models suggest that PBHs could even serve as seeds of the supermassive black holes observed at galactic centers, merging and accreting mass over cosmic time~\cite{Carr:2018rid, Hasinger:2020ptw, Carr:2019kxo, Khlopov:2004sc, Kawasaki:2012kn}.

The aforementioned examples of PBH physics firmly reside in the gravity sector.  Past work to include PBHs in cosmology has focused on inducing a change in the Hubble expansion rate: another application of gravitational physics.  However, PBHs can and will evaporate in the early universe~\cite{Carr:2009jm, Keith:2020jww, Kawasaki:2004qu}. In essence, PBH evaporation converts mass energy density into radiation.  In this work, we will show how the radiation from PBH decay leads to a change in the entropy history of the universe.  A change in entropy can have a profound effect on observable data, namely, the primordial element abundances from Big Bang Nucleosynthesis (BBN). Our work here demonstrates a unique connection between gravitation and nuclear physics.

The outline for the remainder of the paper is as follows. To begin, we give a brief review the early universe conditions conducive to PBH formation and explain how overdense regions collapse into PBHs with masses related to the horizon scale in Sec.\ \ref{sec:formation}. We subsequently calculate the PBH mass fraction at formation, $\beta$, using the Green, Liddle, Malik, and Sasaki (GLMS) framework \cite{Green:2004wb} including a running scalar spectral index, and evolve this abundance to $T=30\,\mathrm{MeV}$ in Sec.\ \ref{sec:massfraction}. We denote the resulting PBH mass fraction by $\beta_{30}(M)$ and use it as an initial condition to the BBN calculations. Sec.\ \ref{sec:bbn_comp} details our implementation of Hawking evaporation into a modified version of the \burst nuclear reaction network code.  We use \burst to evaluate the impact of PBH decay on the synthesis of the light elements \He, \Deu, \he, and \Li. Sec.\ \ref{sec:results} presents our computational results, which reveal a distinct transition near PBH mass $M \approx 10^{10}\,\mathrm{g}$. Two distinct regimes of PBH influence on BBN emerge, and we discuss the individual phenomenologies alongside comoving entropy injection from Hawking radiation. In Sec.\ \ref{sec:conclusion}, we discuss the limitations of our treatment, summarize our findings, and outline their broader implications and possible extensions.

%------------------------------------------------THE FORMATION OF PRIMORDIAL BLACK HOLES------------------------------------------------
\section{\label{sec:formation}THE FORMATION OF PRIMORDIAL BLACK HOLES}
There is a wide variety of models for the formation of PBH's. Below, we focus on a representative model, as an illustration from which we can estimate the mass spectrum, needed for our purposes. The early universe was a hot, dense plasma dominated by relativistic particles, with the radiation energy density determined by the background temperature as $\rho_{ \mathrm{rad} } = \frac{\pi^2}{30} g_{*} T^{4}$, where $g_{*}$ is the relativistic degrees of freedom. From the Friedmann equations, the energy density can also be expressed as $\rho_{\mathrm{rad}} = \frac{3}{8 \pi G 4 t^2} \approx 4.5 \times 10^5 \left( \frac{ 1\,\mathrm{s} }{t} \right)^2\,\mathrm{g}\,\mathrm{cm^{-3}}$. If this radiation density collapsed into a Schwarzschild black hole of radius $r_{\mathrm{s}}$, it would be natural to compare $M_{\mathrm{rad}} = \rho_{\mathrm{rad}} V_{\mathrm{s}}$ with the mass of a Schwarzschild black hole, $M_{\mathrm{s}} = \rho_{\mathrm{s}} V_{\mathrm{s}}$, where $\rho_{\mathrm{s}} =  M_{\mathrm{s}}/\frac{4}{3} \pi r_{\mathrm{s}}^3 \sim 10^{17} \left( M_{\odot}/M_{\mathrm{s}} \right)^2\,\mathrm{g}\,\mathrm{cm^{-3}}$. From $M_{\mathrm{rad}} \gtrsim M_{\mathrm{s}}$, one finds that at cosmic time $t$, the mass of a possible Schwarzschild black hole would scale as~\cite{Carr:2020gox}
\begin{eqnarray} \label{eq:2.1}
    M \gtrsim 10^{10}\,\mathrm{g}\,\frac{t}{10^{-29}\,\mathrm{s}}.
\end{eqnarray}
Coincidentally, this Schwarzschild black hole mass is comparable to the Hubble mass in a radiation-dominated era at cosmic time $t$~\cite{Liddle:2000cg, Keith:2020jww}
\begin{eqnarray} \label{eq:2.2}
    M_{\mathrm{H}} = \frac{G}{2H} \sim 10^{10}\,\mathrm{g}\left( \frac{10^{11}\,\mathrm{GeV}}{T} \right)^2 \left( \frac{106.75}{g_*(T)} \right)^{1/2}.
\end{eqnarray}
The above relation opens up a wide mass range for PBHs potentially formed in the early universe.

However, the fact that we do not observe a PBH zoo around us suggests that additional conditions must be satisfied for PBH formation. A widely studied mechanism proposes that PBHs form from primordial density fluctuations. These fluctuations are characterized by the density contrast, defined as
\begin{eqnarray} \label{eq:density_contrast}
    \Delta(x) \equiv \frac{\rho(x) - \bar{\rho}}{\bar{\rho}},
\end{eqnarray}
where $\rho(x)$ is the local energy density on a space-like hypersurface and $\bar{\rho}$ is the cosmic average at a given instance. Early work by Carr and Hawking~\cite{Carr:1975qj, Carr:1974nx} showed that an overdense region collapses into a PBH if its density contrast exceeds a threshold $\Delta_{\mathrm{min}}$, but remains below an upper bound $\Delta_{\mathrm{max}}$. The lower threshold, which is determined by the Jeans radius, ensures that the region can overcome its own pressure. The upper bound arises because a region with $\Delta > \Delta_{\mathrm{max}}$ has a local geometry of its space-like hypersurface possessing a positive curvature, which is described by a compact and closed Friedmann universe. Unlike regions with $\Delta_{\mathrm{max}} > \Delta > \Delta_{\mathrm{min}}$, which collapse to form PBHs embedded in the background FRW universe, regions with $\Delta > \Delta_{\mathrm{max}}$ evolve as self-contained universes because their geometry is compact and disconnected from the rest of spacetime. 

More recent work has revisited these limits, with~\cite{Harada:2013epa, Harada:2004pe} refining the estimates of $\Delta_{\mathrm{max}}$, and~\cite{Kopp:2010sh} arguing that such estimates depend strongly on the selected geometry, which is not coordinate-invariant and may not impose robust constraints on physical density fluctuations. Therefore, we do not consider the scenario where $\Delta \gg \Delta_{\mathrm{max}}$ and the resulting acausal regions of spacetime.

A more reliable estimate for the lower threshold has been obtained through hydrodynamical simulations~\cite{Evans:1994pj, Niemeyer:1999ak, Hawke:2002rf, Musco:2004ak}. Through the examination of a spherical collapse model in a radiation fluid, these studies suggest that the PBH mass is related to the horizon mass at formation via a critical collapse, $M = K(\Delta - \Delta_{\mathrm{c}})^{\gamma} M_{\mathrm{H}}$, where $K$, $\Delta_{\mathrm{c}}$, and $\gamma$ are constants determined by specific simulation models. A representative set of values, $K = 11.9$, $\gamma = 0.34$, and $\Delta_{\mathrm{c}} = 0.7015$ (Gaussian-curve perturbation), yields $M > 0.1 M_{\mathrm{H}}$ when $\Delta - \Delta_{\mathrm{c}}$ only slightly exceeds $10^{-6}$~\cite{Niemeyer:1999ak}. This framework provides a more accurate connection between density fluctuations and the expected PBH mass. While larger overdensities give rise to more massive PBHs, the lower threshold of the PBH mass can reasonably be identified with the horizon mass given in Eq.~\eqref{eq:2.2}, or at least expected to lie within the same order of magnitude.

%------------------------------------------------PBH mass fraction------------------------------------------------
\section{\label{sec:massfraction}THE PRIMORDIAL BLACK HOLE MASS FRACTION}
Having established that the characteristic PBH mass is comparable to the Hubble mass, we now evaluate the number of PBHs formed at a given epoch. The early universe can be divided into Hubble-sized regions of comoving scale $R = 1/aH$, where primordial inflationary fluctuations generate a density contrast $\Delta(x)$ defined in Eq.~\eqref{eq:density_contrast}. Smoothing $\Delta(x)$ over the scale $R$ with a window function $W$, $\Delta(x,R) = \int_{-\infty}^{\infty} d^3x'W(x-x',R)\Delta(x')$, assigns a single contrast value to each patch. Regions satisfying $\Delta \geq \Delta_{\mathrm{c}}$ collapse into PBHs with masses $M \gtrsim M_{\mathrm{H}}$. 

References~\cite{Green:2004wb, Young:2014ana} estimate the number density of peaks using the statistical approach introduced in~\cite{Bardeen:1985tr}, where the number of peaks follows $n_{\mathrm{peaks}}(\nu > \nu_{\mathrm{c}}) \propto \exp{(-\nu_{\mathrm{c}}^2/2)}$, with $\nu_{\mathrm{c}}^2 \equiv \Delta_{\mathrm{c}}^2 / \sigma^{2}_{\Delta}$. Here, $\Delta_{\mathrm{c}}$ represents the threshold density contrast for collapse, and $\sigma_\Delta^2$ is the variance of the smoothed density field, given by
\begin{eqnarray} \label{eq:3.3}
    \sigma^{2}_{\Delta} &&= \int^{\infty}_{0} \frac{d k}{k} W^{2}(k, R) \mathcal{P}_{\Delta}(k) \\
        &&= \int^{\infty}_{0} \frac{d k}{k} W^{2}(k, R) \frac{4(1+w)^2}{(5+3w)^2} (kR)^{4} \mathcal{P}_{\mathcal{R}}(k).
\end{eqnarray}
Here, $W(k, R)$ is the Fourier transformed window function. $\mathcal{P}_{\Delta}(k)$ and $\mathcal{P}_{\mathcal{R}}(k)$ denote the power spectrum of the smoothed density contrast and the primordial curvature perturbation, respectively. From~\cite{Planck:2018vyg}, the curvature power spectrum is described by $\mathcal{P}_{\mathcal{R}}=A_{\mathcal{R}}\left(k / k_0\right)^{n_s-1}$, where $A_\mathcal{R} \approx 2.1 \times 10^{-9}$ is the amplitude of the scalar power spectrum, $n_s \approx 0.96$ is the scalar spectral index, and $k_0 = 0.05\, \mathrm{Mpc}^{-1}$ is the pivot scale.

Physically, $n_{\mathrm{peaks}}(\nu > \nu_{\mathrm{c}})$ represents the number of Hubble-sized patches that satisfy the collapse condition $\Delta \geq \Delta_{\mathrm{c}}$, corresponding to the number of PBHs with $M \gtrsim M_{\mathrm{H}}$ formed once overdense regions re-enter the horizon. Since the number density depends exponentially on $\Delta_{\mathrm{c}}$, PBHs with masses much larger than $M_{\mathrm{H}}$ are exponentially suppressed. For simplicity in our later analysis, we assume all PBHs form at the time of collapse with masses equal to the Hubble mass.

The PBH mass fraction evaluated at the formation time for PBHs of mass $M$ is defined as $\beta\equiv M \times n_{\mathrm{PBH}}(M)/\rho_{\mathrm{rad}}$, where $\rho_{\mathrm{rad}}$ is the background radiation energy density. It can be expressed as
\begin{eqnarray} \label{eq:3.4}
    \beta = \frac{(n_s + 3)^{3/2}}{6^{3/2} (2 \pi)^{1/2}} (\nu_{\mathrm{c}}^2 -1) e^{- \frac{\nu_{\mathrm{c}}^2}{2}}.
\end{eqnarray}
In~\cite{Green:2004wb}, authors relate $\beta$ to the PBH mass $M \approx M_{\mathrm{H}}$ by connecting the comoving scale $R$ to the horizon mass through $M_{\mathrm{H}} = M_{\mathrm{H,eq}} (k_{\mathrm{eq}} R)^2 \left( g_{*,\mathrm{eq}}/g_{*} \right)^{1/3}$, where $g_{*,\mathrm{eq}} \approx 3$, $k_{\mathrm{eq}} = 0.07\Omega_{\mathrm{m}} h^2\,\mathrm{Mpc}^{-1}$, $M_{\mathrm{H,eq}} = 1.3 \times 10^{49} (\Omega_{\mathrm{m}} h^2)^{-2}\,\mathrm{g}$, and $\Omega_{\mathrm{m}} h^2 = 0.14$. This mapping expresses the PBH mass in terms of comoving scale and cosmological parameters.

\begin{figure}[htbp]
\centering
\includegraphics[width=0.45\textwidth]{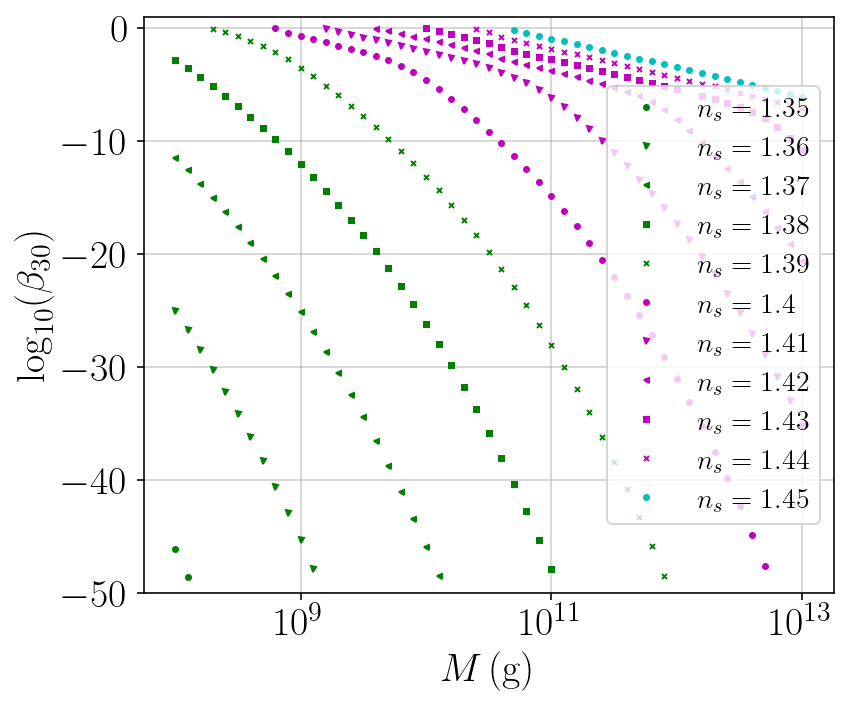}
\vfill
\raisebox{2mm}{% <— tweak this value
  \includegraphics[width=0.45\textwidth]{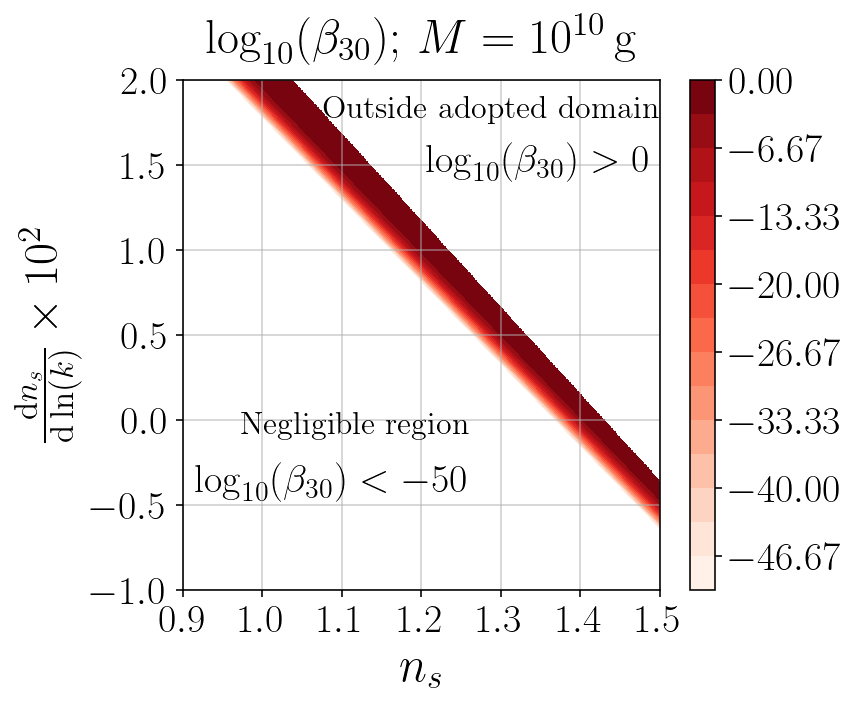}
}
\caption{\label{fig01}\textbf{Top:} PBH mass fraction $\beta_{30}$ at $T=30\,\mathrm{MeV}$, obtained by evolving the formation abundance $\beta(M)$ according to Eqs.~\eqref{eq:3.4} and~\eqref{eq:3.5}. The results are shown for different values of the scalar spectral index $n_s$ and no running. When $n_s<1.38$, the PBH abundance at $30\,\mathrm{MeV}$ remains extremely small ($\log_{10}\beta_{30}\lesssim-10$) within the relevant mass range. \textbf{Bottom:} Contour plot $\log_{10}\beta_{30}(M=10^{10}\,\mathrm{g})$ as a function of $n_s$ and its running $dn_s/d\ln(k)$ at $T = 30\,\mathrm{MeV}$.}
\end{figure}

At horizon entry, PBHs behave as non-relativistic matter with $\rho_{\mathrm{PBH}} \propto a^{-3}$, in contrast to radiation energy density, which redshifts as $a^{-4}$. Consequently, the PBH mass fraction $\beta(M)$ grows linearly with the scale factor $a$. If
PBHs of mass $M$ form at time $t_{\mathrm{i}}$, their abundance at a later time $t_{\mathrm{f}}$ is given by
\begin{eqnarray} \label{eq:3.5}
    \beta(t_{\mathrm{f}},M) = \beta(t_{\mathrm{i}},M) \frac{a(t_{\mathrm{f}})}{a(t_{\mathrm{i}})},
\end{eqnarray}
where the creation time $t_{\mathrm{i}}$ can be estimated using Eq.~\eqref{eq:2.1}. For an adiabatic radiation bath with approximately constant relativistic degrees of freedom, $aT\simeq\mathrm{constant}$, and Eq.~\eqref{eq:3.5} reduces to $\beta(t_{\mathrm{f}},M)\simeq \beta(t_{\mathrm{i}},M)T(t_{\mathrm{i}})/T(t_{\mathrm{f}})$, which allows us to map the PBH abundance at formation to a reference temperature of $30\,\mathrm{MeV}$. We define
\begin{eqnarray} \label{eq:beta30}
    \beta_{30}(M)\equiv\beta(t_{30},M),\qquad T(t_{30})=30\,\mathrm{MeV},
\end{eqnarray}
and use $\beta_{30}(M)$ to label the initial PBH abundance in our BBN calculations. The prescription used to construct this initial condition and to evolve the PBH mass after $t_{30}$ is described in Sec.~\ref{sec:bbn_comp}.

In the top panel of Fig.\ \ref{fig01}, we plot $\log_{10}\beta_{30}$ as a function of the PBH mass $M$. The formation abundance $\beta(M)$ from Eq.~\eqref{eq:3.4} is evolved to $T=30\,\mathrm{MeV}$ using Eq.~\eqref{eq:3.5}. The considered mass range is $10^{8}\,\mathrm{g} < M < 10^{13}\,\mathrm{g}$, since PBHs within this interval can influence BBN, as discussed later in Sec.\ \ref{sec:bbn_comp}. As an example, the first green square indicates that for $n_s = 1.38$, the mass fraction of PBHs with $M = 10^8\,\mathrm{g}$ is $\beta_{30}\approx10^{-2}$. Only models with a scalar spectral index $n_s \geq 1.38$ yield a non-negligible PBH mass fraction at $30\,\mathrm{MeV}$ ($\log_{10}\beta_{30}\gtrsim-10$) within the range relevant to our study. In contrast, the observed primordial power spectrum in the present universe is nearly scale-invariant, with $n_s \sim 1$, which explains why PBH contributions are not expected to be detectable through current BBN constraints.

Recent observations, however, allow for a scale-dependent scalar spectral index, commonly referred to as a \textit{running} spectral index. 
It is parameterized by $n_s(k) = n_s + \frac{1}{2} \frac{\mathrm{d} n_s}{\mathrm{d} \ln (k)} \ln(\frac{k}{k_{0}})$. 
Current measurements~\cite{Planck:2018vyg} give $n_s = 0.9641 \pm 0.0044$ and $\mathrm{d} n_s/\mathrm{d} \ln (k) = -0.0045 \pm 0.0067$, evaluated at the pivot scale $k_0 = 0.05\,\mathrm{Mpc}^{-1}$.

\begin{figure*}[htbp]
\centering
\includegraphics[width=0.90\textwidth]{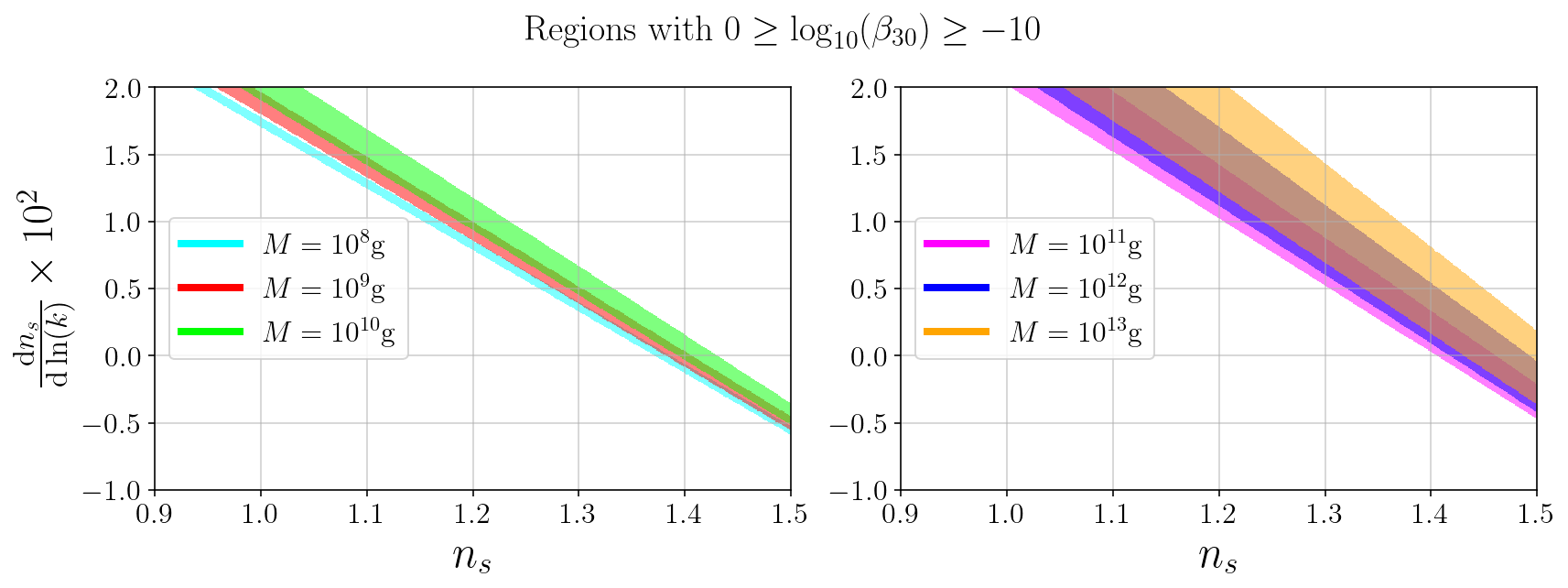}
\caption{\label{fig02} Contours of the PBH mass fraction $\beta_{30}$ for different PBH masses at $T=30\,\mathrm{MeV}$. The region $-10\leq\log_{10}\beta_{30}\leq0$ is the abundance range retained in our parameter scan. As the PBH mass increases, the parameter scan domain broadens and shifts toward higher values of the scalar spectral index $n_s$. Heavier PBHs form more efficiently in spectra with larger $n_s$, consistent with the behavior observed in Fig.\ \ref{fig01}.}
\end{figure*} %$T = 30\,\mathrm{MeV} \approx 3.48 \times 10^{11}\,\mathrm{^{\circ}K}$

We now illustrate how $\beta_{30}(M)$ varies with the scalar spectral index $n_s$ and its running $\mathrm{d}n_s/\mathrm{d}\ln(k)$. As shown in the bottom panel of Fig.\ \ref{fig01}, for $M=10^{10}\,\mathrm{g}$, $\log_{10}\beta_{30}$ exhibits a well-defined pattern in the $[n_s,\mathrm{d}n_s/\mathrm{d}\ln(k)]$ plane. The upper-right region, where $\log_{10}\beta_{30}>0$, lies outside the adopted upper boundary of our scan, while the lower-left region gives a negligible PBH abundance at $30\,\mathrm{MeV}$. Increasing the running broadens the range of $n_s$ included in the scan. Fig.\ \ref{fig02} shows that this trend persists for different PBH masses: as $M$ increases, the parameter scan domain broadens and shifts toward larger $n_s$. These figures identify the primordial-spectrum parameters that produce non-negligible $\beta_{30}$ values for the subsequent BBN calculations.

The introduction of a running spectral index opens the possibility of producing a significant PBH mass fraction even when $n_s$ at CMB scales remains close to unity. To account for this effect in our calculation of $\beta(M)$, we replace the scale-invariant form of the spectrum with its running counterpart. Since no closed analytical expression for the variance in Eq.~\eqref{eq:3.3} exists in this case, we rely on numerical evaluation using the \burst code~\cite{Grohs:2015eua, Grohs:2020xxd}.

The \burst code, originally designed for BBN calculations, has been modified here to incorporate PBHs. A key improvement is its ability to compute a self-consistent scale factor $a(t)$, which is essential for accurately tracking the evolution of the PBH fraction in the universe. We begin by evaluating the PBH mass fraction, $\beta(M)$, at their formation time. This initial $\beta(M)$ depends on the parameters $n_s$, $\mathrm{d} n_s/\mathrm{d}\ln (k)$, and $M$. The obtained mass fraction is then evolved forward to the onset of the BBN era, beginning at a temperature $T = 30\,\mathrm{MeV}$. The resulting value $\beta_{30}(M)$ specifies the initial PBH mass fraction for the BBN calculation.

After determining $\beta_{30}(M)$, we compute the emission of different particle species via Hawking radiation and track how this process alters $\beta(t, M)$ at each time step. The details of this calculation, including how Hawking radiation modifies the PBH mass fraction, will be discussed in Sec.\ \ref{sec:bbn_comp}.

Next, we incorporate the effect of Hawking evaporation on the background energy densities. Specifically, the radiation and matter energy densities are modified by including contributions from PBHs. Using Friedmann’s equations, 
\begin{eqnarray} \label{eq:3.6}
    H(t)^2 = \frac{8 \pi G}{3} \left[  \frac{\rho_{\mathrm{PBH}(t_{\mathrm{i}})}  a^3(t_{\mathrm{i}})}{a^3(t)}  + \frac{\rho_{\mathrm{rad}(t_{\mathrm{i}})}  a^4(t_{\mathrm{i}})}{a^4(t)} \right],
\end{eqnarray}
we then evolve the scale factor $a(t)$. Because PBHs behave as non-relativistic matter, the ratio of $a(t)$ at the present time $t_0$ to its value at BBN time $t_{\mathrm{BBN}}$ is greater than in the standard $\Lambda$CDM model without PBHs. To ensure consistency, if $a(t_0)$ differs from unity by more than $10^{-6}$, we iteratively adjust the initial estimate of $a(t_{\mathrm{BBN}})$ until convergence is achieved.

%------------------------------------------------CHANGE IN BBN FROM PBH DECAY------------------------------------------------
\section{\label{sec:bbn_comp}CHANGE IN BBN FROM PBH DECAY}
We have established both the physical conditions under which PBHs can form and the statistical framework that determines their mass fraction $\beta_{30}$ at the starting temperature of our BBN calculation. With these ingredients in hand, we now turn to the central question of how the presence of PBHs modifies the dynamics of BBN. By emitting energetic particles through Hawking radiation, PBHs can inject entropy and alter reaction rates in the early plasma~\cite{Vainer:1977, Vainer:1978, Boccia:2024nly}, thereby modifying the abundances of light elements relative to standard BBN predictions. Before addressing these modifications, it is useful to briefly recall the main features of standard BBN in the absence of PBHs.

In the standard scenario, BBN is modeled within a homogeneous and isotropic $\Lambda$CDM universe, together with the Standard Model of particle physics that includes three neutrino species. After inflation, the universe entered a radiation-dominated epoch during which reactions among elementary particles remained in equilibrium. As the universe cools to $T \approx 0.8 \,\mathrm{MeV}$, the weak interaction rate governing neutron--proton conversions falls below the Hubble expansion rate, satisfying $\Gamma_{\mathrm{W}} < H$ and marking the freeze-out of the neutron-to-proton ratio~\cite{Cyburt:2015mya, Bernstein:1988ad, Mukhanov:2003xs}.

\begin{table}[htbp]
\centering
\begin{tabular}{|c|c|c|}
\hline
$\text{n} + e^{+} \rightleftharpoons \text{p} + \bar{\nu_{e}}$ & $\text{n} + \nu_{e} \rightleftharpoons \text{p} + e^{-}$ & $\text{n} \rightleftharpoons \text{p} + e^{-} + \bar{\nu_{e}}$ \\
\hline
\end{tabular}
\caption{Weak reactions inter-converting neutrons to protons.\label{tab:01}}
\end{table}

It is important to note, however, that the notion of a sharp ``freeze-out'' is only an approximation. Even after the weak interaction rate drops below the Hubble rate, neutron decay continues and lepton capture reactions still modify the neutron--proton ratio. As the universe cools further to $T \approx 0.1\,\mathrm{MeV}$, the photon background becomes dilute enough for deuterium to survive photodisintegration, despite its binding energy of $2.2\,\mathrm{MeV}$. The small number of baryons per photon ($\eta \equiv n_{\mathrm{b}}/n_\gamma$) delays this process until relatively low temperatures, but once it occurs, light nuclei can form and persist.
%Nuclear statistical equilibrium (NSE) initially governs the abundances, but as the expansion proceeds and photodisintegration becomes inefficient, synthesis continues out of dissociation.

Because of the Coulomb barrier, the synthesis of nuclei beyond $\mathrm{A} = 8$ is strongly suppressed, leaving only trace amounts of heavier elements (e.g., $^9\mathrm{Be}$, $^{10}\mathrm{B}$). As a result, the observable relics of BBN are $^{1}\mathrm{H}$, \Deu, \he, \He, also with $^6\mathrm{Li}$ and \Li produced at much lower levels. The primordial abundance of $^6\mathrm{Li}$ remains uncertain because it is readily destroyed in stellar environments, whereas \Li is consistently overproduced in standard BBN compared with observational data --- a long-standing discrepancy known as the ``lithium problem''~\cite{Jedamzik:2007cp, Asplund:2005yt, Iocco:2012vg}. Comparisons between the observed and predicted primordial abundances of these light elements provide valuable insights into the physics of the early universe. In this work, we will focus on \He, \Deu, \he, and \Li.

In standard BBN, nearly all free neutrons are incorporated into \He. Consequently, the primordial \He abundance is determined primarily by the neutron-to-proton ratio prior to nucleosynthesis. This ratio can be approximated by the Boltzmann factor $\np \approx e^{-\Delta m / T_{\mathrm{f}}} \approx 1/5$, where $\Delta m = m_n - m_p$ and $T_{\mathrm{f}} \approx 0.8 \,\mathrm{MeV}$ is the freeze-out temperature~\cite{Kolb:1990vq}. Below this temperature, weak interactions involving free neutron decay and lepton capture processes continue to reduce the ratio to $\np \approx 1/7$, yielding a helium mass fraction of $\yp = n(\He) m(\He) / n_{\mathrm{b}} m_{\mathrm{b}} = 2(\np)/[1+(\np)] \approx 0.25$. Observational determinations of \He from emission lines in H-II regions suggest $\yp = 0.2345 \pm 0.0026$~\cite{Peimbert:2000yj}, while more recent analyses using Markov Chain Monte Carlo techniques yield $\yp = 0.2449 \pm 0.0040$~\cite{Aver:2015iza}.

Although the freeze-out estimate provides useful intuition, it is only approximate and neglects important nuclear processes. Given the precision of modern observations of $\yp$, reliable predictions require following the full weak and nuclear reaction network rather than relying on such simplified arguments. This is particularly relevant for understanding the nontrivial variations of $\yp$ that emerge in our analysis of PBH effects in Sec.\ \ref{sec:5.1}.

Among the light elements, deuterium (\Deu) is particularly valuable as a probe of the baryon-to-photon ratio, since the deuterium-to-hydrogen ratio ($\Deu / \mathrm{H}$) decreases monotonically with $\eta$ and is highly sensitive to it~\cite{Tytler:2000qf, Iocco:2008va}. A higher baryon density allows nuclear reactions to proceed more efficiently, synthesizing nearly all of the deuterium into helium. Moreover, D/H can be directly measured from absorption spectra in metal-poor, damped Lyman-$\alpha$ systems. Recent observations report $\Deu / \mathrm{H} = (2.527 \pm 0.030) \times 10^{-5}$~\cite{Cooke:2017cwo}. \he provides another sensitive indicator of $\eta$, though its interpretation is complicated by continuous production and destruction in stellar environments~\cite{Kawasaki:2017bqm}. As a result, present-day measurements of \he are less robust indicators of $\eta$ than deuterium. Measurements in Galactic H-II regions yield $\he / \mathrm{H} = (1.1 \pm 0.2) \times 10^{-5}$~\cite{Bania:2002yj}. \Li remains the most puzzling case: standard BBN, calibrated by the baryon density inferred from CMB data, predicts abundances significantly higher than those observed in metal-poor halo stars, where $\Li / \mathrm{H} = (1.23^{+0.68}_{-0.32}) \times 10^{-10}$~\cite{Ryan:1999rxn,Fields:2011zzb}.

Having outlined the key features of standard BBN, we now turn to how this picture is modified in the presence of PBHs. PBHs can influence BBN in several ways, but one of the crucial effects arises from Hawking radiation, through which they inject high-energy particles into the primordial plasma. A black hole of mass $M$ radiates particles at a characteristic temperature~\cite{Hawking:1975vcx}
\begin{eqnarray} \label{eq:4.1}
    T_{\mathrm{BH}} = \frac{1}{8 \pi G M} \approx 10^{6} \left(10^{10}\,\mathrm{g}/M\right) \,\mathrm{MeV}.
\end{eqnarray}
Any Standard Model particle with energy below this temperature can be emitted. While the precise energy of an emitted particle depends on its species, it is sufficient for our purposes to approximate the emission energy as $E \approx T_{\mathrm{BH}}$.

Hawking radiation causes the black hole to lose mass at a rate~\cite{MacGibbon:1991tj}
\begin{eqnarray} \label{eq:4.2}
    \frac{\mathrm{d}M}{\mathrm{d}t} = -5.34\times 10^{25} f(M) \left(1\,\mathrm{g}/M\right)^{2} \,\mathrm{g}\,\mathrm{sec^{-1}},
\end{eqnarray}
where $f(M)$ encodes the number of particle species emitted and is normalized to unity when $M \gg 10^{17}\,\mathrm{g}$. This formulation, introduced in studies of quark--gluon jet emission, provides the basis for estimating PBH evaporation timescales.

Integrating this expression yields the black hole lifetime,
\begin{eqnarray} \label{eq:4.3}
    \tau_{\mathrm{BH}} &&\approx 6.24 \times 10^{-27} (M/1\,\mathrm{g})^3 f(M)^{-1}\,\mathrm{sec}\nonumber\\
        &&\approx 1.98 \times 10^{-34} (M/1\,\mathrm{g})^3 f(M)^{-1}\,\mathrm{yr}.
\end{eqnarray}

For example, with $M = 10^{13}\,\mathrm{g}$ and $f(M) \approx 12.5$, the lifetime is $\tau_{\mathrm{BH}} \sim 5 \times 10^{11}\,\mathrm{s} \approx 1.6 \times 10^{4}\,\mathrm{yr}$. A PBH of mass $10^{15}\,\mathrm{g}$ instead survives for $\tau_{\mathrm{BH}} \sim 10^{10}\,\mathrm{yr}$, comparable to the current age of the universe.

The function $f(M)$ accounts for the evaporation rates of different particle species, including leptons $l$, quarks $q$, neutrinos $a$, gluons $g$, $W$ and $Z$ bosons, as well as the Higgs boson. Its explicit form, incorporating spin-dependent weights and degrees of freedom, is given by~\cite{Lunardini:2019zob}
\begin{widetext}
\begin{eqnarray} \label{eq:4.4}
    f(M) = &&2 f_{1} + 4 f_{1/2}^{1} \left[ \sum_{l = e,\mu,\tau} \exp\left(- \frac{M}{s_{1/2} M_{l} }\right) + 3\sum_{q} \exp\left(- \frac{M}{s_{1/2} M_{q} }\right) \right]
    + 2 \eta_{\nu}^{N} f_{1/2}^{0} \sum_{a = 1,2,3} \exp\left(- \frac{M}{s_{1/2} M_{a} }\right) \nonumber\\
    &&+ 16 f_{1} \exp\left(- \frac{M}{s_{1} M_{g} }\right)
    + 3 f_{1} \left[ 2 \exp\left(- \frac{M}{s_{1} M_{W} }\right) + \exp\left(- \frac{M}{s_{1} M_{Z} }\right) \right] + f_{0} \exp\left(- \frac{M}{s_{0} M_{\mathrm{Higgs}} }\right),
\end{eqnarray}
\end{widetext}
where $M_{j} = 1/(8 \pi G m_{j})$ corresponds to the PBH mass whose Hawking temperature equals the rest mass $m_{j}$ of a given particle species. Massless photons contribute to the first term, $2 f_{1}$, while massless gluons are assigned an effective mass $m_{g} = 0.6\,\mathrm{GeV}$, corresponding to the infrared cutoff in QCD renormalization~\cite{Halzen:1984mc}. The coefficients $s_{s}, f_{s}^{q}$, and $\eta_{\nu}^{N}$ encode the spin and charge dependent parameters, and distinguish between Dirac and Majorana neutrinos. Their numerical values are summarized in
\begin{eqnarray} \label{eq:4.5}
    &&s_{0} = 2.66,\,s_{1/2} = 4.53,\,s_{1} = 6.04 \nonumber\\
    &&f_{0} = 0.267,\,f_{1} = 0.060 \nonumber\\
    &&f_{1/2}^{0} = 0.147~\text{(uncharged)},\,f_{1/2}^{1} = 0.142~\text{(charged)} \nonumber\\
    &&\eta_{\nu}^{\mathrm{Dirac}} =2,\eta_{\nu}^{\mathrm{Majorana}} = 1.
\end{eqnarray}

While PBHs emit the full spectrum of Standard Model particles, not all of them propagate freely into the cosmological plasma. Whether an emitted particle significantly affects the background depends on its mean free path relative to the black hole's characteristic and inter-separation scale. This criterion is key to identifying which emissions influence BBN and may lead to inhomogeneous heating during BBN. Previous studies~\cite{Vainer:1977} examined PBHs with masses in the range $10^{9}\,\mathrm{g} \leq M \leq 10^{12}\,\mathrm{g}$, and found that only photons, with mean free path $l_{\gamma} \sim 10^{-9}\,\mathrm{cm}$, and neutrinos, with $l_{\nu} \sim 10^{-11}\,\mathrm{cm}$ travel farther than the Schwarzschild radius of a PBH with $M \sim 10^{10}\,\mathrm{g}$ ($r_{\mathrm{s}} \sim 10^{-17}\,\mathrm{cm}$). The vicinity of a black hole would therefore be transparent only to photons and neutrinos, which can directly participate in background nuclear reactions (see Tab.\ \ref{tab:01} and\ \ref{tab:02}).

For heavier particles, the mean free path is much smaller, and therefore annihilate quickly around a cloud of PBH's horizon size~\cite{Vainer:1977}. Within our approximation, ignoring inhomogeneous heating during BBN is justified. Previous works~\cite{Carr:1976zz, He:2022wwy} have therefore suggested that evaporating PBHs do not release such species as free-streaming particles. Instead, their emissions thermalize within an optically thick ``hot envelope'' surrounding the PBHs. High-energy particles undergo repeated scatterings with the background plasma, quickly reaching a new thermal equilibrium. A similar treatment of rapid thermalization of PBH emissions is used in~\cite{Boccia:2024nly}. Following this picture, we assume that massive particles influence BBN only indirectly: by depositing their energy into the surrounding medium, they heat the plasma and contribute to its entropy density. While we do not consider the direct interaction of injected hadrons/mesons with the background plasma, complementary analyses based on different assumptions instead track the hadronic/meson component explicitly and study its effect on the neutron–to-proton conversion rate~\cite{Wu:2025ovd}.
%It is worth noting that most previous studies of PBH effects on BBN assumed conservation of the comoving entropy density in a radiation-dominated universe. In our framework, however, this assumption can break down when the PBH mass fraction approaches values comparable to unity.

In the \burst calculations, we first scan a parameter space spanned by the scalar spectral index $n_s$, its running $\mathrm{d} n_s/\mathrm{d}\ln (k)$, and the PBH mass $M$. We retain models whose PBH mass fraction at the starting temperature satisfies $\beta_{30}(n_s, \mathrm{d}n_s/\mathrm{d}\ln (k), M) \leq 1$. This is the adopted upper boundary of our parameter scan domain. It is not an observational bound and does not by itself guarantee radiation domination. The scalar spectral index is varied within $0.9 \leq n_s \leq 1.5$, while the running is constrained to $-0.010 \leq \mathrm{d}n_s/\mathrm{d}\ln (k) \leq 0.020$. For the PBH mass, we consider the range $10^{8}\,\mathrm{g} \leq M \leq 10^{13}\,\mathrm{g}$. The lower bound reflects the fact that PBHs lighter than $10^{8}\,\mathrm{g}$ evaporate completely before BBN, while the upper bound corresponds to $\tau \approx 5 \times 10^{11}\,\mathrm{s}$, such that evaporation ends well before observables like the CMB are affected.

For each retained combination of $n_s$, $\mathrm{d}n_s/\mathrm{d}\ln(k)$, and $M$, the corresponding $\beta_{30}$ is incorporated into the BBN reaction network. In our framework, PBHs primarily act as sources emitting neutrinos and plasma particles through Hawking radiation. At high temperatures, before photons and neutrinos decouple (when neutrino--electron and neutrino--positron weak interactions remain efficient), all species share a common temperature governed by local thermal equilibrium. As the universe cools and photons and neutrinos decouple (at an equivalent of $T_{\mathrm{dec}} \approx 1.5\,\mathrm{MeV}$ in the absence of PBHs~\cite{Dodelson:2020bqr}), the neutrino and plasma sectors evolve separately.  We make the assumption that neutrinos remain in Fermi--Dirac equilibrium with zero chemical potential, thereby neglecting energy down-scattering transport processes. Reference~\cite{Grohs:2020xxd} gives the prescription for evolving the temperature under such constraints. Meanwhile, the plasma particles continue to thermalize through Compton scattering, double Compton scattering, and bremsstrahlung. The non-neutrino energy density emitted by PBHs are assumed to instantaneously thermalize with the surrounding plasma, injecting entropy into the background, while the emitted neutrinos directly participate in background weak and nuclear reactions. %whereas photons and the heavier species increase the background plasma temperature indirectly through entropy injection.

\begin{table*}[htbp]
\centering
\begin{tabular}{|c|c|c|c|}
\hline
${}^{1}_{1}\mathrm{H}$(n, $\gamma$)${}^{2}_{1}\mathrm{H}$ & ${}^{2}_{1}\mathrm{H}$(n, $\gamma$)${}^{3}_{1}\mathrm{H}$ & ${}^{3}_{2}\mathrm{He}$(n, $\gamma$)${}^{4}_{2}\mathrm{He}$ & ${}^{6}_{3}\mathrm{Li}$(n, $\gamma$)${}^{7}_{3}\mathrm{Li}$ \\
\hline
${}^{3}_{2}\mathrm{He}$(n, p)${}^{3}_{1}\mathrm{H}$ & ${}^{7}_{4}\mathrm{Be}$(n, p)${}^{7}_{3}\mathrm{Li}$ & ${}^{6}_{3}\mathrm{Li}$(n, $\alpha$)${}^{3}_{1}\mathrm{H}$ & ${}^{7}_{4}\mathrm{Be}$(n, $\alpha$)${}^{4}_{2}\mathrm{He}$ \\
\hline
${}^{2}_{1}\mathrm{H}$(p, $\gamma$)${}^{3}_{2}\mathrm{He}$ & ${}^{3}_{1}\mathrm{H}$(p, $\gamma$)${}^{4}_{2}\mathrm{He}$ & ${}^{6}_{3}\mathrm{Li}$(p, $\gamma$)${}^{7}_{4}\mathrm{Be}$ & ${}^{6}_{3}\mathrm{Li}$(p, $\alpha$)${}^{3}_{2}\mathrm{He}$ \\
\hline
${}^{7}_{3}\mathrm{Li}$(p, $\alpha$)${}^{4}_{2}\mathrm{He}$ & ${}^{2}_{1}\mathrm{H}$($\alpha$, $\gamma$)${}^{6}_{3}\mathrm{Li}$ & ${}^{3}_{1}\mathrm{H}$($\alpha$, $\gamma$)${}^{7}_{3}\mathrm{Li}$ & ${}^{3}_{2}\mathrm{He}$($\alpha$, $\gamma$)${}^{7}_{4}\mathrm{Be}$ \\
\hline
${}^{2}_{1}\mathrm{H}$(${}^{2}_{1}\mathrm{H}$, n)${}^{3}_{2}\mathrm{He}$ & ${}^{2}_{1}\mathrm{H}$(${}^{2}_{1}\mathrm{H}$, p)${}^{3}_{1}\mathrm{H}$ & ${}^{3}_{1}\mathrm{H}$(${}^{2}_{1}\mathrm{H}$, n)${}^{4}_{2}\mathrm{He}$ & ${}^{3}_{2}\mathrm{He}$(${}^{2}_{1}\mathrm{H}$, p)${}^{4}_{2}\mathrm{He}$ \\
\hline
${}^{3}_{2}\mathrm{He}$(${}^{3}_{2}\mathrm{He}$, 2p)${}^{4}_{2}\mathrm{He}$ & ${}^{7}_{3}\mathrm{Li}$(${}^{2}_{1}\mathrm{H}$, n$\alpha$)${}^{4}_{2}\mathrm{He}$ & ${}^{7}_{4}\mathrm{Be}$(${}^{2}_{1}\mathrm{H}$, p$\alpha$)${}^{4}_{2}\mathrm{He}$ &  \\
\hline
\end{tabular}
\caption{Nuclear reaction chain in the \burst code~\cite{Grohs:2015eua, Grohs:2020xxd}.\label{tab:02}}
\end{table*}

Because PBHs increase the comoving entropy density through Hawking radiation, we adjust the initial baryon-to-photon ratio $\eta$ at $T = 30\,\mathrm{MeV}$ to a greater value than in standard BBN. This adjustment ensures that the final baryon-to-photon ratio at $T = 8.6 \times 10^{-4}\,\mathrm{keV} \approx 10^{4}\,\mathrm{K}$ matches the observed CMB value of $\eta = 6.1 \times 10^{-10}$ \cite{Planck:2018vyg}. Practically, we determine a self-consistent $\eta$ by an iterative correction procedure analogous to that used for the scale factor $a(t)$ in Sec.\ \ref{sec:massfraction}: we evolve the system with an initial guess for $\eta$, compare the resulting value with the target, and repeat until the relative difference falls below $10^{-6}$. This yields consistent values of $a(t)$ and $\eta$ that fully incorporate both the PBH mass fraction and entropy injection from Hawking radiation.  

With the initial PBH mass fraction $\beta_{30}$ and the corresponding background initial conditions, we evolve the full nuclear reaction network forward in time, including all processes listed in Tables \ref{tab:01} and\ \ref{tab:02}. At each timestep, the PBH mass-loss rate is computed using Eq.~\eqref{eq:4.2} and Eq.~\eqref{eq:4.4}, and the associated energy transfer to the plasma is tracked as a continuous heating source. PBH photons and neutrinos, owing to their long mean free paths, deposit their energy directly into the corresponding plasma, while heavier PBH decay products heat the plasma indirectly through secondary interactions confined in the hot envelope. This energy injection modifies the background photon and neutrino temperatures, as well as the neutrino number density, thereby influencing both the weak and nuclear reaction rates and the Hubble expansion rate. In this way, PBH evaporation affects the thermal history and the nuclear processes of BBN, providing the physical basis for the results discussed in the next section.

%------------------------------------------------Results------------------------------------------------
\section{RESULTS} \label{sec:results}

\subsection{Entropy and Primordial Abundances} \label{ssec:entropy_bbn}

\begin{figure}[htbp]
\centering
\includegraphics[width=0.45\textwidth]{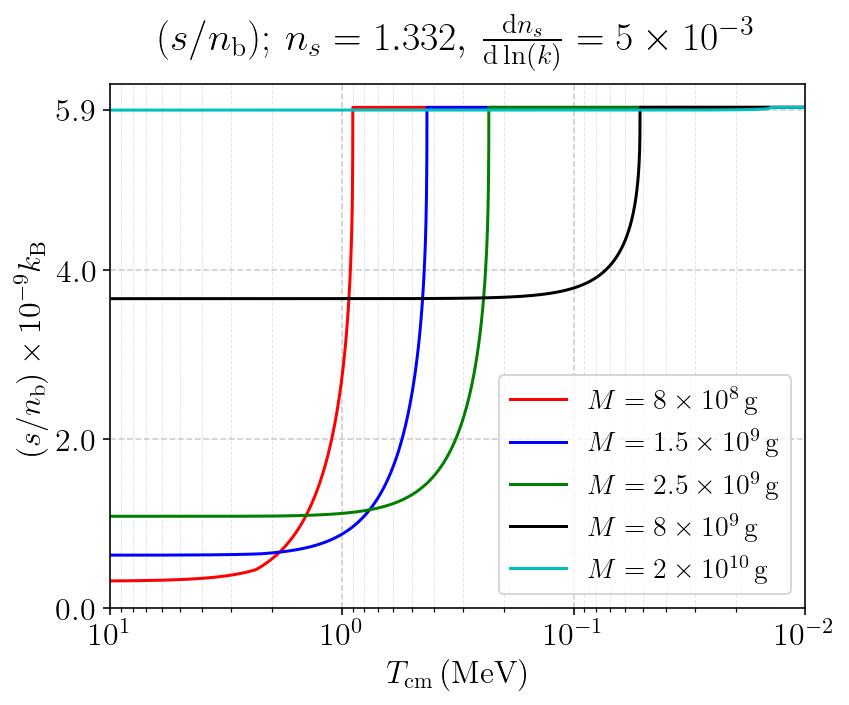}
\caption{\label{fig03}Evolution of the comoving entropy density per baryon, $(s/n_{\mathrm{b}})$ as a function of comoving temperature quantity, $T_{\mathrm{cm}}$, in the case $n_s = 1.3$ and $\mathrm{d} n_s/\mathrm{d} \ln(k) = 5\times10^{-3}$. Each curve represents a different PBH mass and therefore a different initial value of $\beta_{30}$.}
\end{figure}

Here we illustrate how energy injection from PBH evaporation increases the comoving entropy density as the universe expands and cools through the nucleosynthesis epoch. Previous studies generally considered very small formation abundances, often quoted as $\beta\sim10^{-20}$~\cite{Carr:2009jm, Acharya:2020jbv, Keith:2020jww}, for which the energy released through evaporation produced negligible plasma heating. In contrast, our parameter scan extends to $\beta_{30}\sim1$, where PBH evaporation significantly alters the comoving entropy density, making it essential to account for this variation to ensure a self-consistent cosmological evolution in our PBH-BBN simulations.

Figure\ \ref{fig03} shows the evolution of the comoving entropy density per baryon ${s/n_{\mathrm{b}}}$ as a function of decreasing comoving temperature.  We consider a single value of the spectral index and the running, although the results in this figure generalize to other $[n_s,\mathrm{d} n_s/\mathrm{d} \ln(k)]$ pairs.  Each curve represents a different PBH mass.
For models with negligible PBH evaporation (cyan curve with $M = 2 \times 10^{10}\,\mathrm{g}$), the evolution nearly reproduces the standard cosmology: the comoving baryon number and entropy densities remain conserved after baryogenesis, yielding the present-day values $(s/ n_{\mathrm{b}} ) = 5.9 \times 10^9 k_{\mathrm{B}}$. When PBH evaporation becomes significant, entropy injection from Hawking radiation increases $( s/ n_{\mathrm{b}} )$. To reproduce the observed CMB value at late times, the universe must begin with a lower initial entropy density. The discontinuity in the slope of each curve marks the event of complete PBH evaporation corresponding to its mass.

\begin{figure*}[htbp]
\centering
\includegraphics[width=1\textwidth]{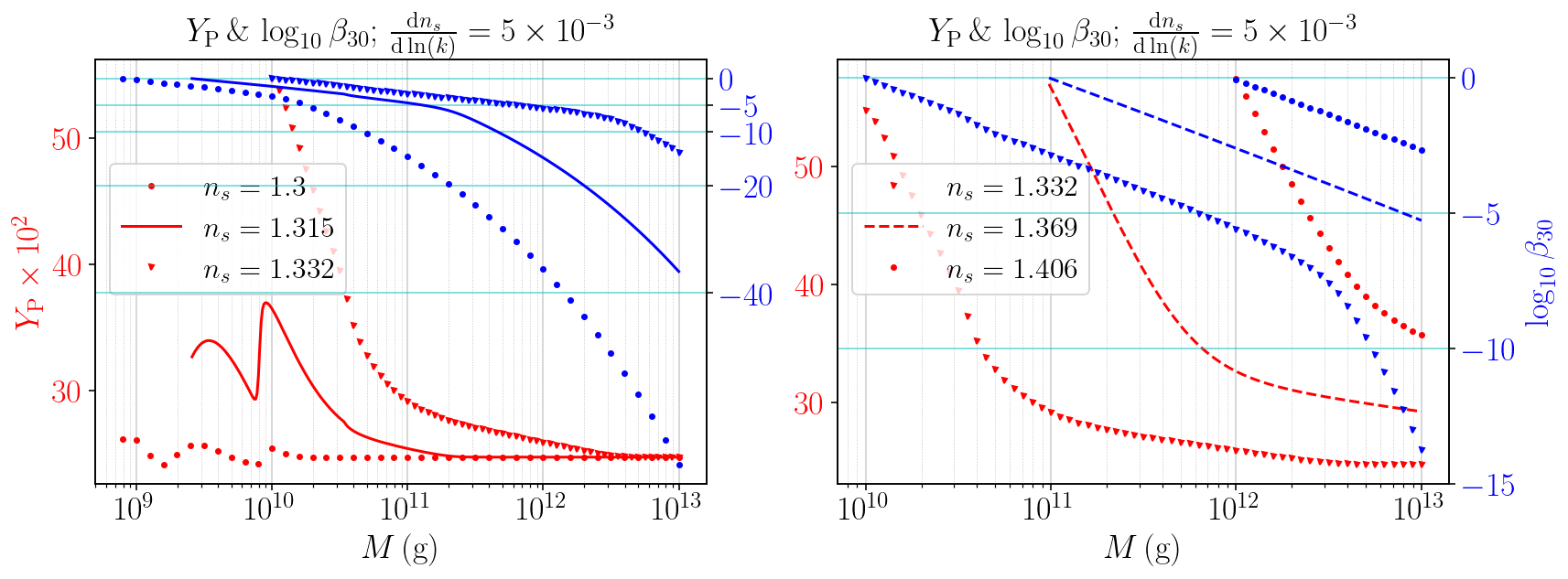}
\caption{\label{fig04}Final \He mass fraction \yp obtained from the \burst BBN network calculation (red, left axis) and the logarithm of the initial PBH mass fraction, $\log_{10}\beta_{30}$, specified at the starting temperature $T=30\,\mathrm{MeV}$ (blue, right axis), as functions of PBH mass $M$. Results are shown for several values of $n_s$ at fixed $\mathrm{d}n_s/\mathrm{d}\ln(k)=5\times10^{-3}$.}
\end{figure*}

\begin{figure*}[htbp]
\centering
\includegraphics[width=1\textwidth]{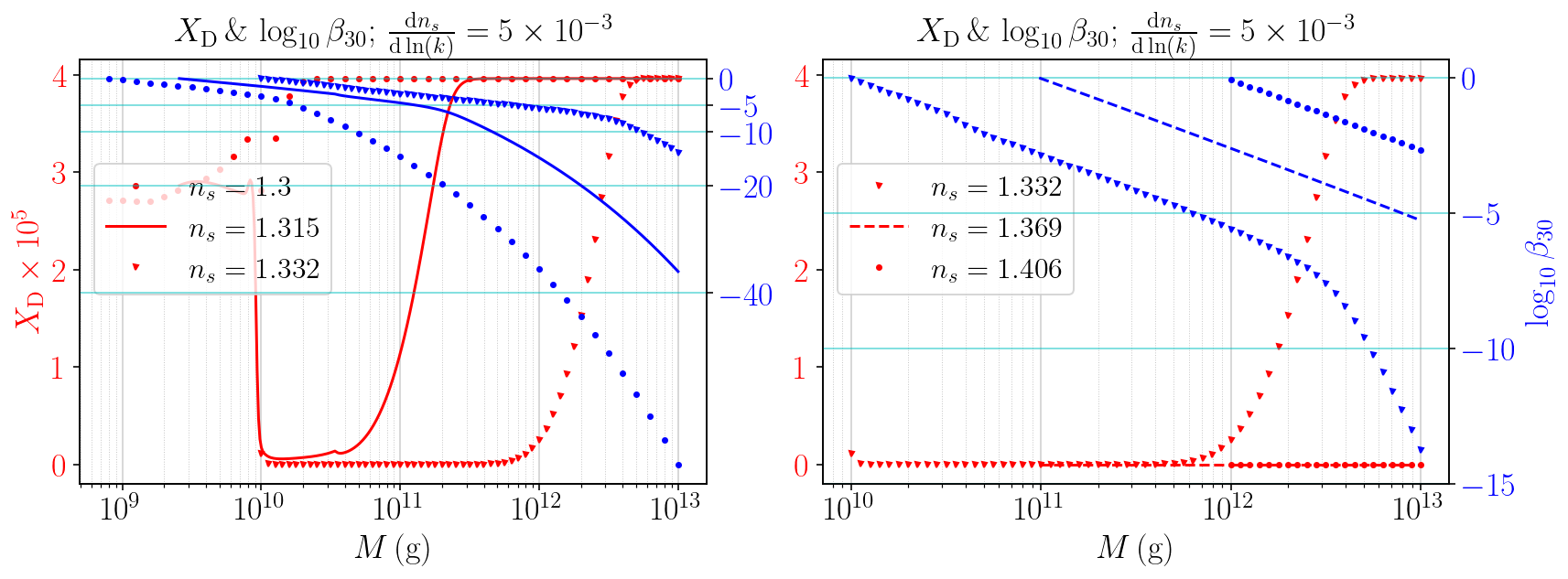}
\caption{\label{fig05}Same as Fig.\ \ref{fig04} but for the deuterium mass fraction \xd.}
\end{figure*}

We note that the lower-mass models ($M < 10^{10}\,\mathrm{g}$) in Fig.\ \ref{fig03} completely evaporate when the background comoving temperature $T_{\mathrm{cm}}$ falls between $1\,\mathrm{MeV}$ and $0.1\,\mathrm{MeV}$. This temperature range is particularly important as it coincides with the epoch when light-element abundances depart from nuclear statistical equilibrium as the nuclear reaction rates fall below the Hubble expansion rate.  Consequently, the $M \sim 10^{10} \,\mathrm{g}$ value presents a transition region where complete PBH evaporation occurs before/after out-of-equilibrium nucleosynthesis.

We give the primordial abundance results for our $[n_s, \mathrm{d} n_s/ \mathrm{d} \ln (k), M]$ extended PBH-BBN models. Figures\ \ref{fig04}--\ \ref{fig07} show the primordial abundances \yp, \xd, \xhe, and \xli, respectively, as red symbols over a range of PBH masses and scalar spectral indices. The right panels focus on masses above the $M\sim10^{10}\,\mathrm{g}$ threshold, whereas the left panels include masses below the threshold. The corresponding values of $\beta_{30}$ are shown as blue symbols.
As an example, the first red triangle in both panels of Fig.\ \ref{fig04} corresponds to $n_s=1.332$, $\mathrm{d}n_s/\mathrm{d}\ln(k)=5\times10^{-3}$, and $M=10^{10}\,\mathrm{g}$. This model has $\beta_{30}\approx1$ at the start of the calculation and yields a final \He mass fraction $\yp\approx0.55$.

\begin{figure*}[htbp]
\centering
\includegraphics[width=1\textwidth]{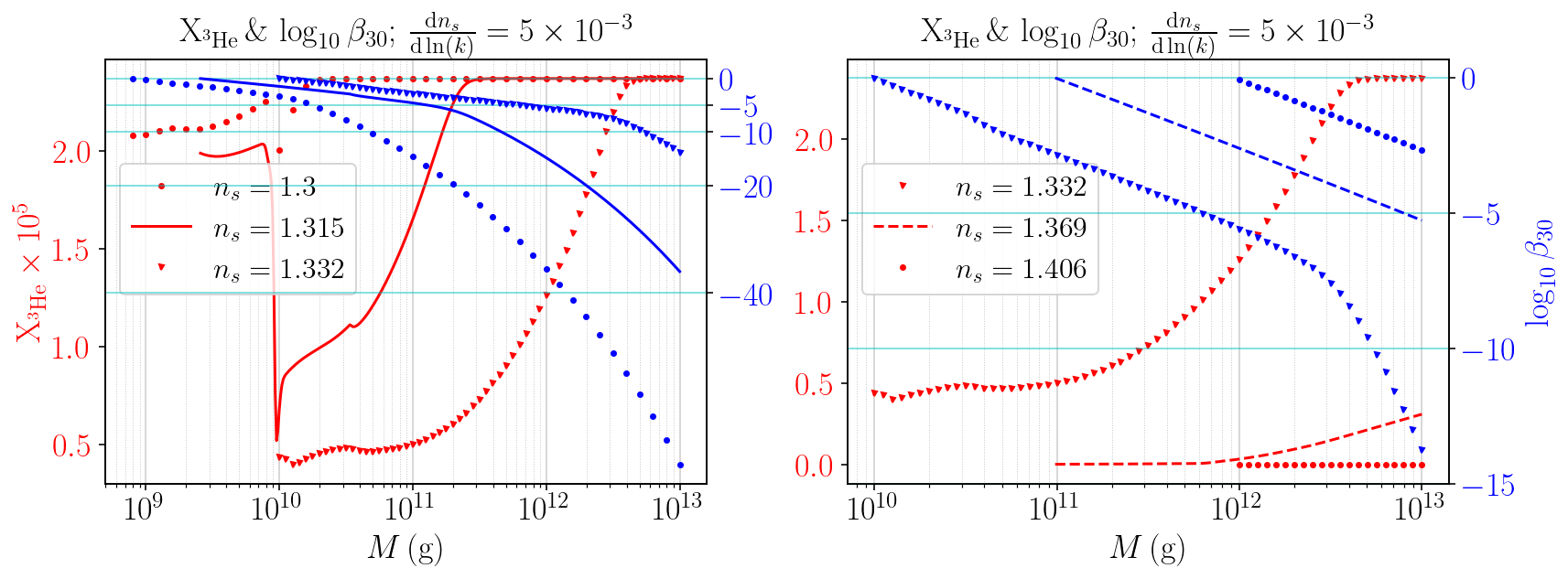}
\caption{\label{fig06}Same as Fig.\ \ref{fig04} but for the $\he$ mass fraction \xhe.}
\end{figure*}

\begin{figure*}[htbp]
\centering
\includegraphics[width=1\textwidth]{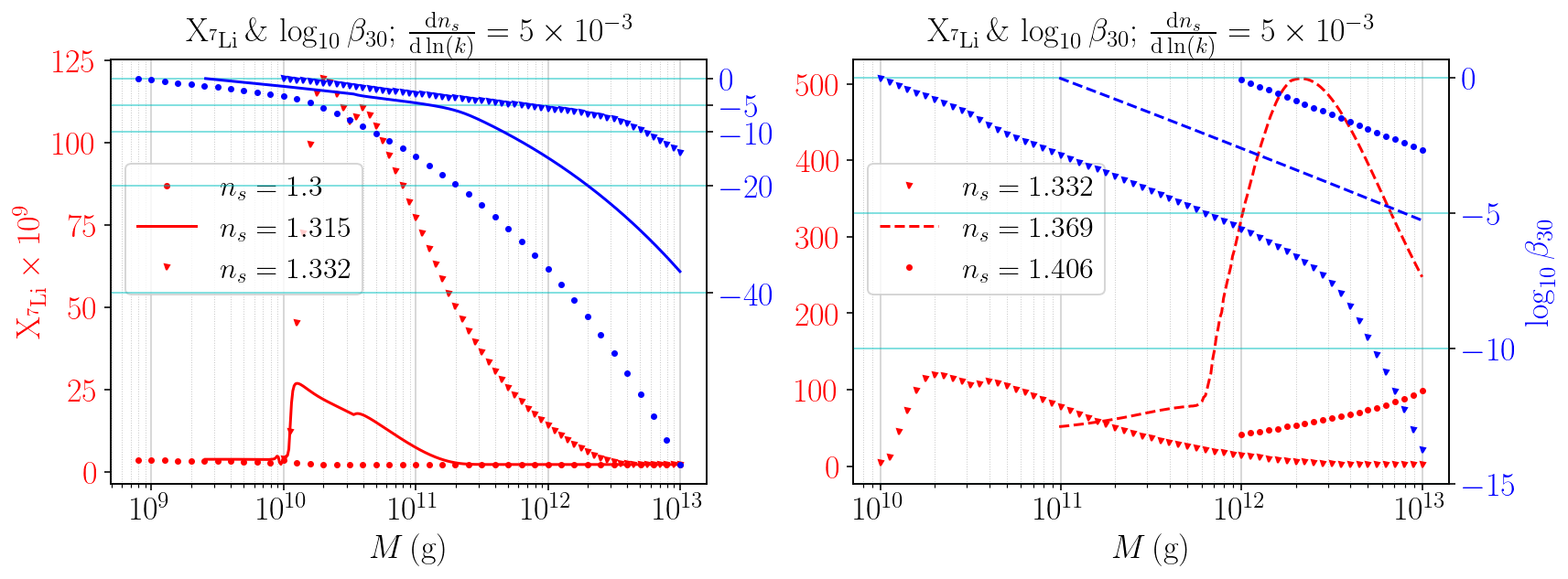}
\caption{\label{fig07}Same as Fig.\ \ref{fig04} but for the \Li mass fraction \xli.}
\end{figure*}

In each plot from Fig.\ \ref{fig04} to \ref{fig07}, we observe that when the PBH mass fraction satisfies $\beta_{30} < 10^{-10}$, the presence of PBHs has a negligible impact on the light element abundances because of their very small fraction. In this regime, the abundances converge to values consistent with a standard computation of BBN absent PBHs.  Those predicted primordial abundances are consistent with the observed values, except for $\xli$, which remains roughly a factor of 5 higher than the observed value~\cite{Jedamzik:2007cp, Asplund:2005yt}.

These four figures reveal a clear change in behavior of light element abundances around $M \approx 10^{10}\,\mathrm{g}$, indicating that PBHs with $M < 10^{10}\,\mathrm{g}$ and those with $M > 10^{10}\,\mathrm{g}$ influence BBN through different mechanisms. For $M < 10^{10}\,\mathrm{g}$, $\yp$ shows an oscillatory dependence on PBH mass, while the deuterium abundance, $\xd$, and the helium-3 abundance, $\xhe$, are slightly suppressed in the presence of PBHs. In this region, the lithium-7 abundance, $\xli$, experiences a modest enhancement.  

For $M > 10^{10}\,\mathrm{g}$, $\yp$ increases monotonically as a function of $\beta_{30}$, largely independent of PBH mass. Both $\xd$ and $\xhe$ are strongly suppressed at large $\beta_{30}$, regardless of $M$. In contrast, $\xli$ does not follow a similar pattern: its maximum variation does not occur at the largest $\beta_{30}$. Instead, the peak in the $n_s = 1.332$ curve appears at $\log_{10}(\beta_{30}) \approx -2$, while the peak in the $n_s = 1.369$ curve occurs at $\log_{10}(\beta_{30}) \approx -4$. Overall, in the region $M > 10^{10}\,\mathrm{g}$, the light element abundances are more strongly correlated with the PBH mass fraction $\beta_{30}$ than with the PBH mass $M$.

The net contribution of PBHs with mass $M$ to light element abundances is complex. Nevertheless, it can be decomposed into three main components: (i) high-energy photons from Hawking radiation, (ii) high-energy neutrinos from Hawking radiation, and (iii) modification of the Hubble rate due to the presence of PBHs.

The injection of high-energy photons through Hawking radiation heats the background photons, which are tightly coupled to baryons before decoupling. As a result, the average kinetic energy of all particle species in thermal equilibrium increases. The increased energy of charged particles raises the tunneling probability through the Coulomb barrier, further amplifying reaction rates. On the other hand, a hotter photon bath strengthens photodisintegration processes, which counteract nuclear synthesis.  The larger net rates result in keeping the light-element abundances closer to nuclear statistical equilibrium~\cite{Burbidge:1957vc}.

Similarly, the injection of high-energy neutrinos via Hawking radiation heats the background neutrino bath, enhancing reaction rates involving neutrinos. An increased neutrino number density also shifts the reactions inter-converting protons and neutrons (see Tab.\ \ref{tab:01}) to a new weak equilibrium, thereby modifying the neutron-to-proton ratio $\np$.  

Finally, if a significant fraction of PBHs remains during BBN, they substantially increase the Hubble rate $H(t)$. A higher expansion rate shortens the effective duration of BBN, precipitating an earlier nuclear freeze-out when reaction rates drop below the Hubble rate.

%------------------------Section 5.2------------------------
\subsection{Distinct BBN Behavior Across the Threshold} \label{sec:5.1}
To determine the origin of the $M \sim 10^{10}\,\mathrm{g}$ threshold for the primordial abundance patterns in Figs.\ \ref{fig04} -- \ref{fig07}, we will consider an argument based on weak freeze-out, i.e., the period of BBN when the processes in Tab.\ \ref{tab:01} fall below the Hubble expansion rate.  Although this is a protracted epoch, we will utilize an instantaneous ``freeze-out temperature'' \tfo for simplicity.  How \tfo changes across models with different $\beta_{30}$ will allow us to focus on the relationship between energy density and the primordial abundances while temporarily ignoring changes from the injection of entropy.

\begin{figure}[htbp]
\centering
\includegraphics[width=0.45\textwidth]{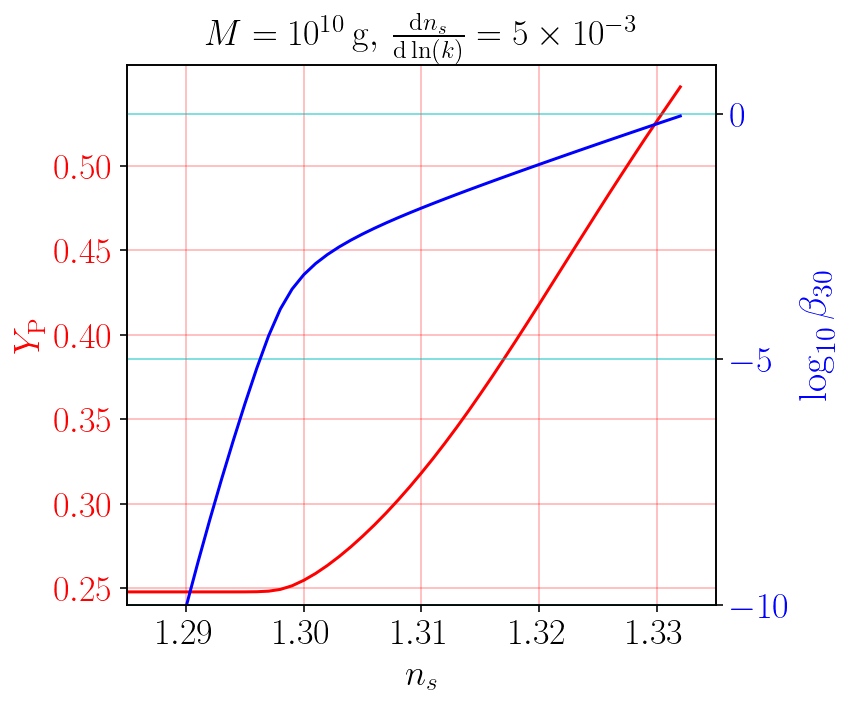}
\caption{\label{fig08}\yp (red) and $\log_{10}\beta_{30}$ (blue) plotted against $n_s$ for $\mathrm{d} n_s/\mathrm{d}\ln(k)=5\times10^{-3}$.}
\end{figure}

To wit, we will suppose a typical weak interaction rate scales as $\Gamma_{\mathrm{w}} \sim T^5$. In the PBH-dominated limit, and at fixed reference temperature, the matter density scales as $\rho_{\mathrm{PBH}}\propto\beta_{30}T^3$, so that $H\propto\beta_{30}^{1/2}T^{3/2}$. Equating the two rates at freeze-out yields
\begin{eqnarray} \label{freeze_out_T}
    \tfo \sim \beta_{30}^{1/7}.
\end{eqnarray}
As $\beta_{30}$ increases, weak freeze-out occurs earlier, leading to larger \np and hence larger \yp. Figure\ \ref{fig08} shows this qualitative trend for $M=10^{10}\,\mathrm{g}$. We increase $\beta_{30}$ (blue curve) by increasing the scalar spectral index while holding its running fixed. As $\beta_{30}$ grows from a negligible value to order unity, the PBH energy density becomes comparable to the radiation density and \yp (red curve) increases from its baseline value to above $0.5$. We surmise that the freeze-out approximation is consistent with the changes in \yp for $M\gtrsim10^{10}\,\mathrm{g}$.

\begin{figure}[htbp]
\centering
\includegraphics[width=0.45\textwidth]{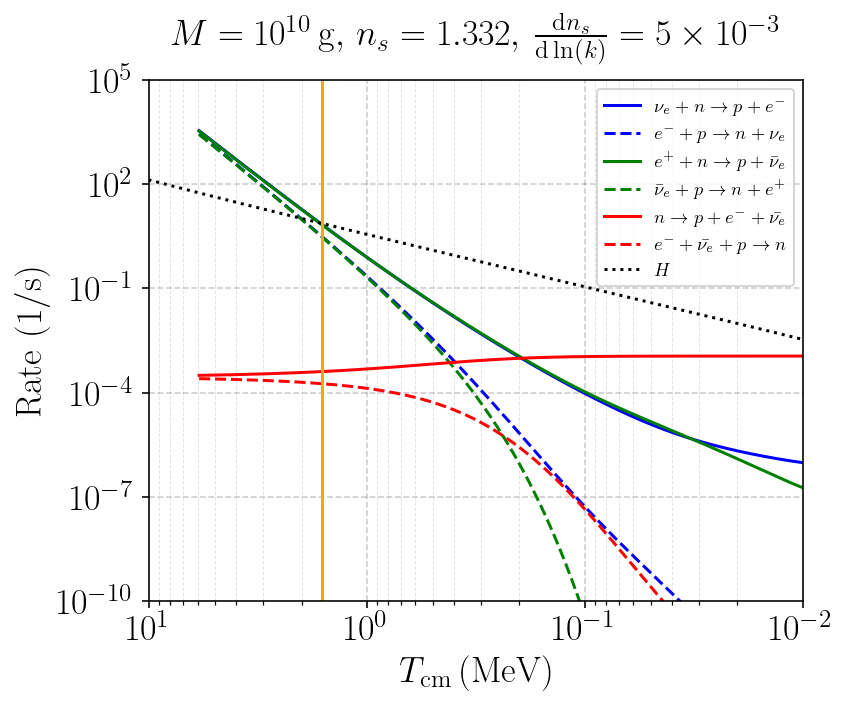}
\vfill
\hspace*{2.5mm}
\includegraphics[width=0.48\textwidth]{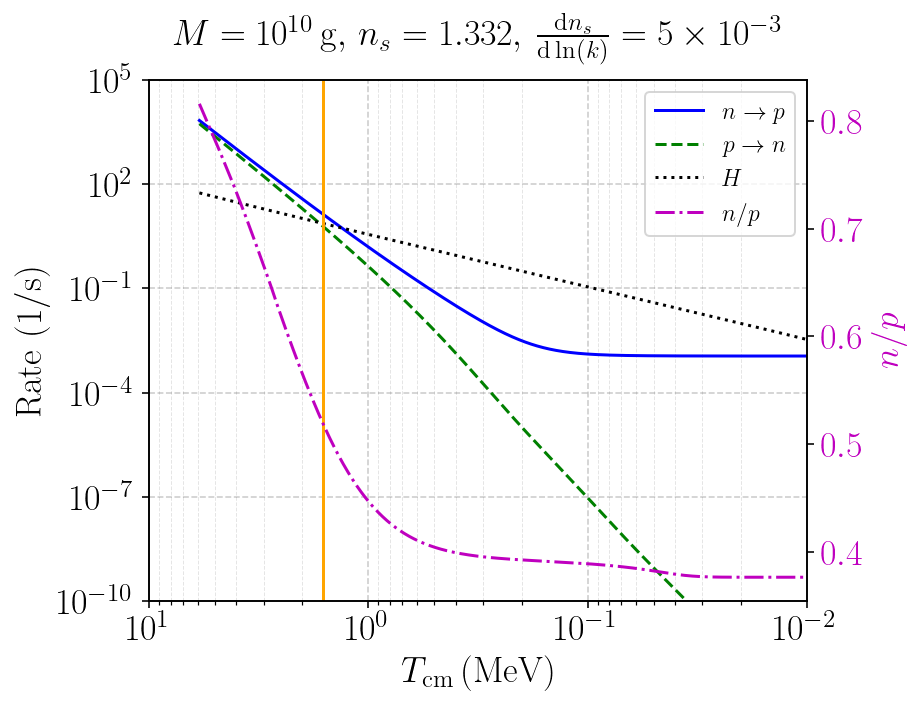}
\hspace*{-2.5mm}
\caption{\label{fig09}\textbf{Top:} Reaction rates from Table \ref{tab:01} and the Hubble expansion rate as functions of the comoving temperature $T_{\mathrm{cm}}$ for a model where $M=10^{10}\,\mathrm{g}$, $n_s=1.332$, and $\mathrm{d} n_s/\mathrm{d}\ln(k)=5\times10^{-3}$.  The orange vertical line gives the epoch when the net neutron destruction rate is equal to the Hubble expansion rate. 
\textbf{Bottom:} All forward reactions ($n \rightarrow p$ in left panel) summed into the solid blue line, and all backward reactions ($p \rightarrow n$) summed into the green dashed line. Also included is the Hubble expansion rate, and $n/p$.}
\end{figure}

\begin{figure*}[htbp]
\centering
\begin{minipage}{0.48\textwidth}
    \centering
    \begin{picture}(0,0)
        \put(-121,-20){\textbf{(a)}} % adjust numbers for position
    \end{picture}
    \includegraphics[width=\linewidth]{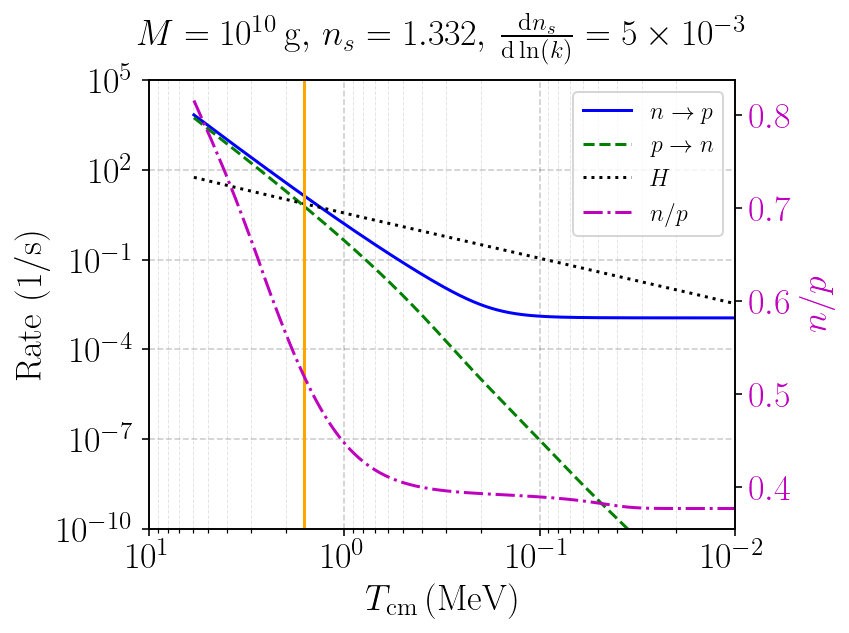}
\end{minipage}
\hfill
\begin{minipage}{0.48\textwidth}
    \centering
    \begin{picture}(0,0)
        \put(-121,-20){\textbf{(b)}}
    \end{picture}
    \includegraphics[width=\linewidth]{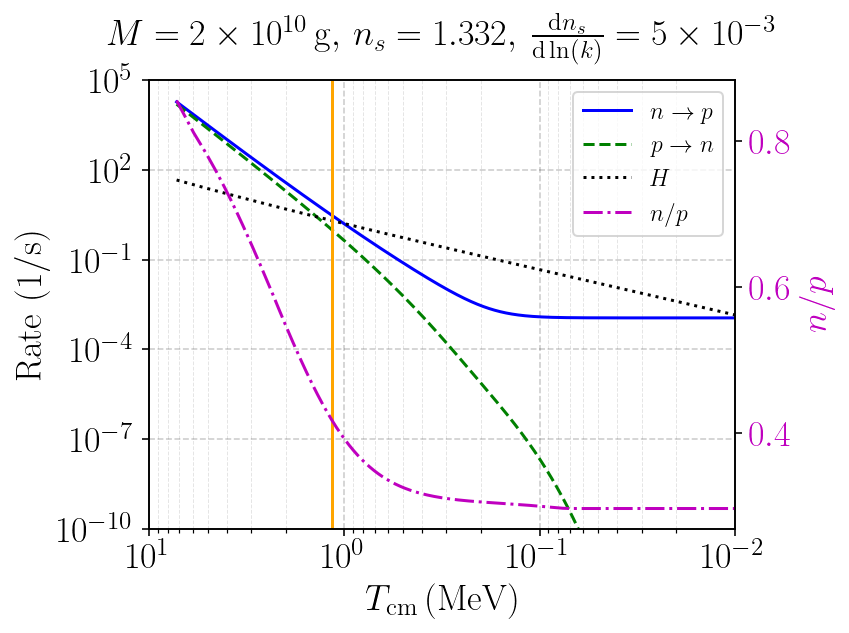}
\end{minipage}

\medskip

\begin{minipage}{0.48\textwidth}
    \centering
    \begin{picture}(0,0)
        \put(-121,-20){\textbf{(c)}}
    \end{picture}
    \includegraphics[width=\linewidth]{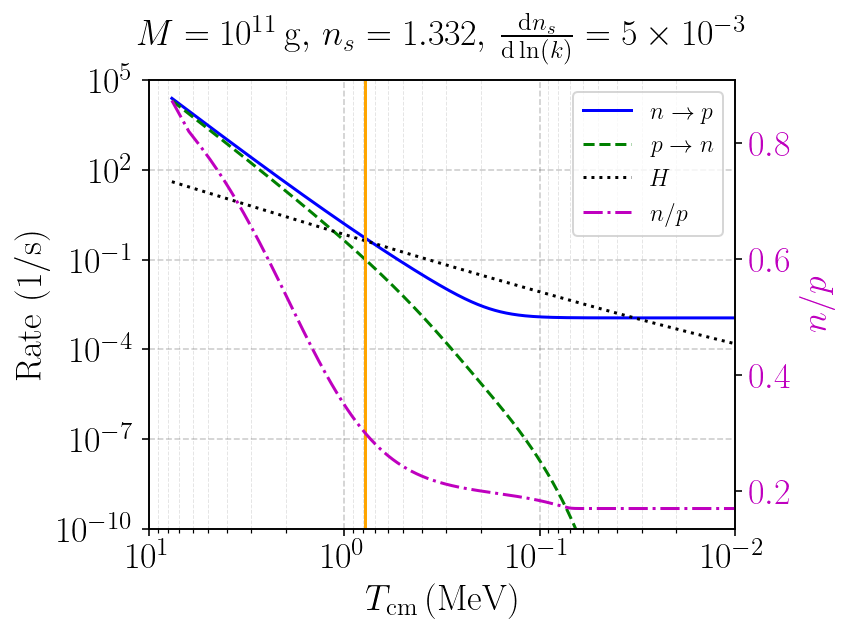}
\end{minipage}
\hfill
\begin{minipage}{0.48\textwidth}
    \centering
    \begin{picture}(0,0)
        \put(-121,-20){\textbf{(d)}}
    \end{picture}
    \includegraphics[width=\linewidth]{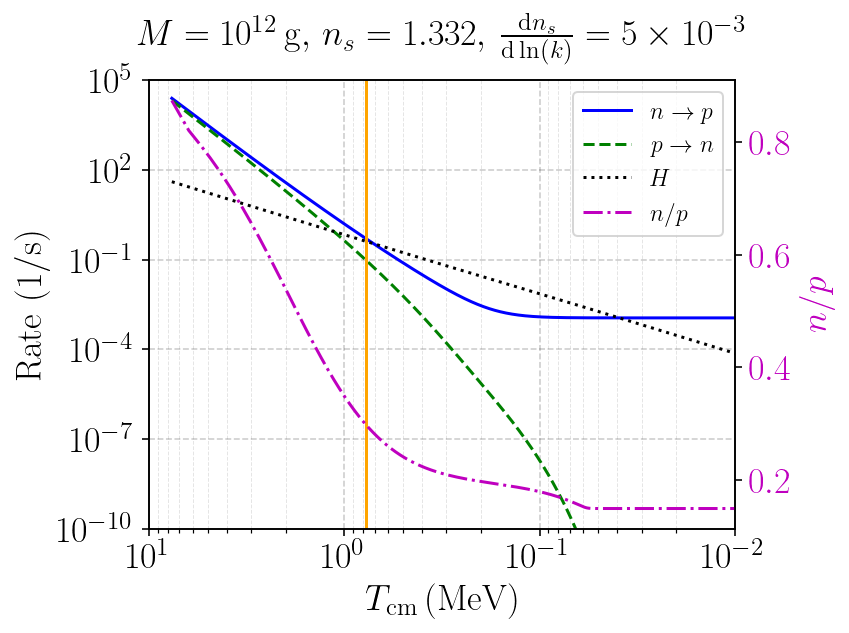}
\end{minipage}

\caption{\label{fig10}Same as the the bottom panel of Fig.\ \ref{fig09} for a model where $n_s=1.332$, and $\mathrm{d} n_s/\mathrm{d}\ln(k)=5\times10^{-3}$.  Each panel corresponds to a different mass, ranging from $10^{10}$ to $10^{12}\,\mathrm{g}$.}
\end{figure*}

Before we consider the mass range $M \lesssim 10^{10}\,\mathrm{g}$, we give more details on the neutron-to-proton interconversion rates in the presence of PBHs.  In Fig.\ \ref{fig09}, we plot neutron-to-proton reaction rates, the Hubble expansion rate, and the neutron-to-proton ratio.  The top panel gives the three neutron destruction rates (solid lines), three neutron creation rates (dashed lines), and the Hubble expansion rate for $M=10^{10}\,\mathrm{g}$, $n_s=1.332$, and $\mathrm{d} n_s/\mathrm{d}\ln(k)=5\times10^{-3}$.  As seen in Fig.\ \ref{fig04}, these parameter values correspond to a model with $\beta_{30}\simeq1$.  This pushes \tfo (orange vertical line) to an earlier epoch.  The bottom panel of Fig.\ \ref{fig09} includes the evolution of \np, along with summed forward and reverse neutron-to-proton rates and the Hubble expansion rate.  The early epoch of weak freeze-out leads to the larger asymptotic value of \np.

Figure\ \ref{fig10} includes the bottom panel of Fig.\ \ref{fig09} as panel (a), together with three models having $M>10^{10}\,\mathrm{g}$. As the mass increases from panels (a) to (d), $\beta_{30}$ decreases, producing the lower Hubble rates shown by the dotted black curves. The smaller $\beta_{30}$ also shifts \tfo to lower temperatures. For $M=10^{12}\,\mathrm{g}$, $\beta_{30}\simeq10^{-6}$ and the evolution of \np approaches the standard BBN result. Figures \ref{fig08} and \ref{fig10} therefore support the conclusion that the high-mass abundance trends are controlled primarily by the change in the Hubble expansion rate.

\begin{figure*}[htbp]
\centering
\begin{minipage}{0.48\textwidth}
    \centering
    \begin{picture}(0,0)
        \put(-121,-20){\textbf{(a)}} % adjust numbers for position
    \end{picture}
    \includegraphics[width=\linewidth]{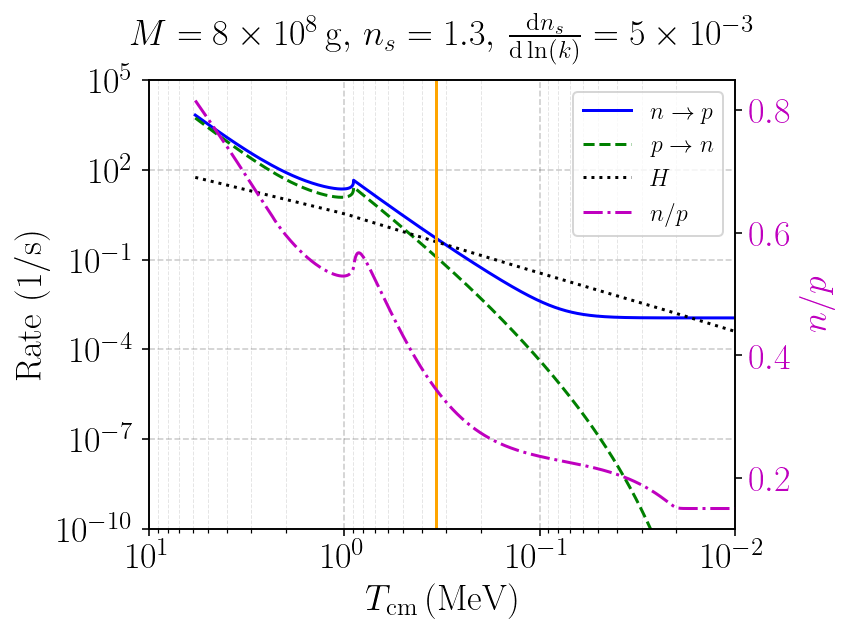}
\end{minipage}
\hfill
\begin{minipage}{0.48\textwidth}
    \centering
    \begin{picture}(0,0)
        \put(-121,-20){\textbf{(b)}}
    \end{picture}
    \includegraphics[width=\linewidth]{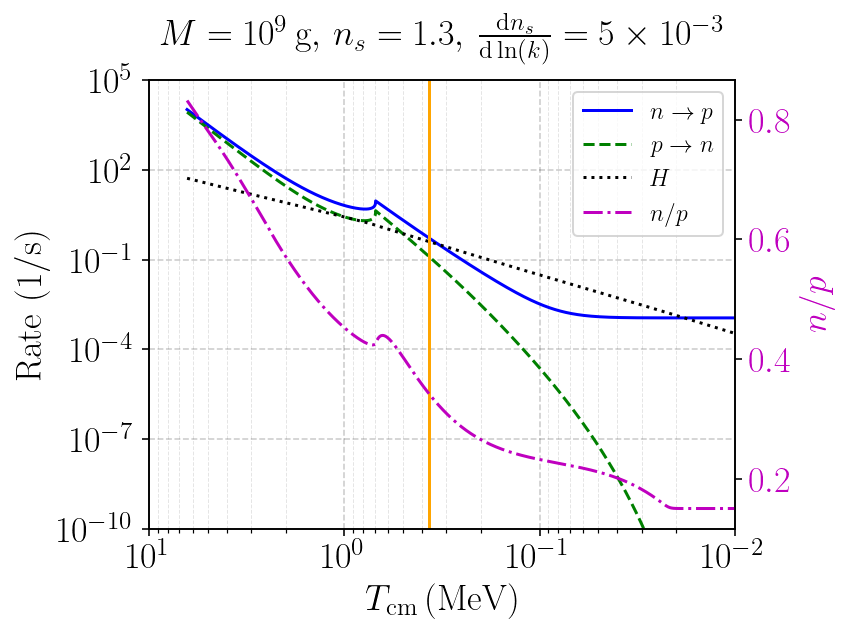}
\end{minipage}

\medskip

\begin{minipage}{0.48\textwidth}
    \centering
    \begin{picture}(0,0)
        \put(-121,-20){\textbf{(c)}}
    \end{picture}
    \includegraphics[width=\linewidth]{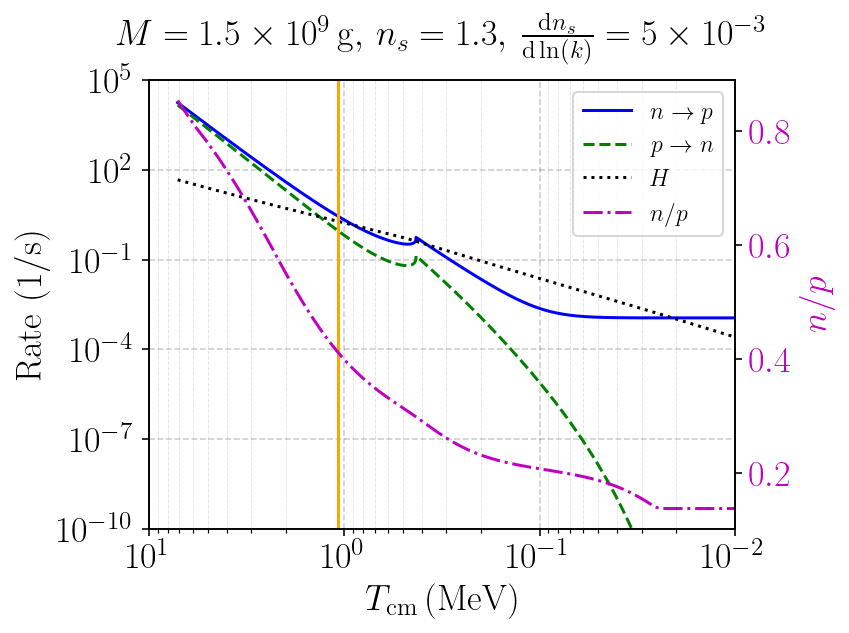}
\end{minipage}
\hfill
\begin{minipage}{0.48\textwidth}
    \centering
    \begin{picture}(0,0)
        \put(-121,-20){\textbf{(d)}}
    \end{picture}
    \includegraphics[width=\linewidth]{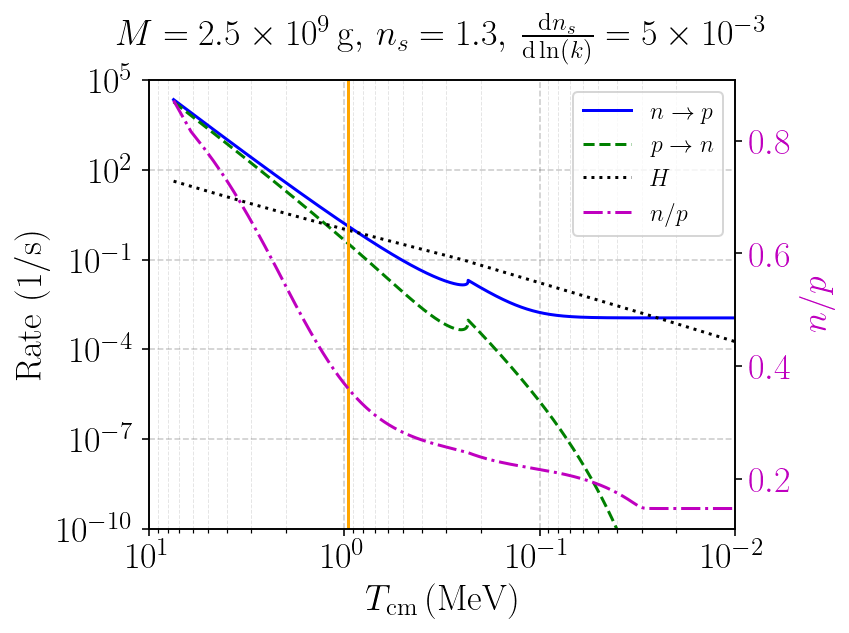}
\end{minipage}

\caption{\label{fig11}Same as Fig.\ \ref{fig10} for a model where $n_s=1.3$, and $\mathrm{d} n_s/\mathrm{d}\ln(k)=5\times10^{-3}$.  Each panel corresponds to a different mass, ranging from $8\times10^{8}$ to $2.5\times10^{9}\,\mathrm{g}$.}
\end{figure*}

As emphasized in Sec.\ \ref{ssec:entropy_bbn}, PBHs with $M < 10^{10}\,\mathrm{g}$ can completely evaporate during BBN, altering the thermal and nuclear dynamics. This behavior is illustrated in Fig.\ \ref{fig11}, which shows how PBH evaporation modifies the evolution of the reaction rates and the Hubble rate. As in Fig.\ \ref{fig11}, $\beta_{30}$ decreases with increasing mass in panels (a) through (d).  The cusps readily apparent in the neutron-to-proton rates and \np (yet clearly absent in $H$) reflect the epoch when the PBHs cease to exist.  Heavier PBHs possess longer lifetimes, so evaporation cusps appear at progressively lower temperatures.
However, we notice that \tfo does not follow the same monotonic trend as Fig.\ \ref{fig10}, with \tfo increasing by nearly a factor of 3 between panels (b) and (c).  Here, the entropy injection is of paramount importance.  The relationship between \np and increasing entropy cannot be so easily dissected for these lower masses.

The conventional neutron freeze-out picture, in which a higher $\np$ ratio always yields a larger $\yp$, breaks down in the low-mass regime. For example, panel (a) of Fig.\ \ref{fig11} gives $\np < 0.4$ at the freeze-out temperature, while panel (c) yields $\np \gtrsim 0.4$. However, panel (a) corresponds to the first peak in the $\yp$ curve of Fig.\ \ref{fig04}, which is higher than the first valley associated with panel (c). This complex behavior arises from the interplay between the PBH lifetime, which sets the timing of evaporation, and $\beta_{30}$, which sets the initial normalization of the PBH energy density. Their subsequent evolution determines whether evaporation can momentarily enhance the reaction rates relative to $H$. As a result, for $M \leq 10^{10}\,\mathrm{g}$, evaporation induces an oscillatory pattern in $\yp$, indicating that a more detailed treatment is needed to generalize this analysis to other light elements.

%------------------------------------------------THERMALIZATION OF PBH EMISSIONS------------------------------------------------
\section{Discussion and Summary} \label{sec:conclusion}
\subsection{Thermalization of PBH Emissions} \label{sec:thermalization}
In our calculation, we assume that particles emitted from PBHs thermalize instantaneously and spread homogeneously into the background plasma. This assumption allows us to include the energy injection from Hawking radiation in a homogeneous BBN reaction network. In reality, the emitted particles first lose their energy near each PBH, and the deposited energy then spreads through the plasma. We briefly discuss both processes below and estimate when our assumption is reasonable.

The thermalization of high-energy particles emitted by a black hole has been studied in~\cite{Das:2021wei,He:2022wwy,He:2024wvt}. In a weakly coupled relativistic gauge plasma, the energy loss of a particle with momentum $p \gg T$ is dominated by nearly collinear inelastic splittings. Although destructive interference suppresses these splittings through the Landau--Pomeranchuk--Migdal (LPM) effect, their fractional energy-loss rate exceeds the elastic contribution by a factor of order $\sqrt{p/T}$~\cite{He:2022wwy}. Using the notation of~\cite{He:2024wvt}, the LPM splitting rate and the thermalization time are
\begin{eqnarray} \label{eq:thermal_lpm_v3}
    \Gamma_{\mathrm{LPM}}(k,T)
    &&=c_{\mathrm{LPM}}\alpha^2T
    \sqrt{\frac{T}{k}},
    \qquad \nonumber\\
    t_{\mathrm{th}}
    &&=\Gamma_{\mathrm{LPM}}^{-1}(T_s,T),
\end{eqnarray}
where $k$ is the energy of a daughter particle, $\alpha$ is the relevant coupling, and $c_{\mathrm{LPM}}$ is a numerical coefficient. The source temperature is $T_s=T_{\mathrm{BH}}$ for an evaporating black hole, $T$ is the temperature of the background plasma, and the primary particles have $p\sim T_s$. The splitting with $k\simeq p/2$ transfers the largest amount of energy and takes the longest time within the cascade. It therefore determines $t_{\mathrm{th}}$, while the lower-energy daughters split more rapidly until their energies become comparable to $T$.

In these calculations, particles with energies of order $T_{\mathrm{BH}}$ are assumed to join the local thermal plasma. The initial energy is deposited near $r_{\mathrm{LPM}}\simeq t_{\mathrm{th}}$, producing a hot shell around the black hole. The thermal energy then diffuses into the surrounding plasma. Authors in~\cite{He:2024wvt} characterize the distance reached after a time $\tau$ by
\begin{eqnarray} \label{eq:thermal_diff_v3}
    r_d(\tau)
    \sim
    \sqrt{\frac{\lambda_{\mathrm{mfp}}(r_d)\,\tau}{3}},
\end{eqnarray}
where $\lambda_{\mathrm{mfp}}$ is the mean free path of the high-energy particle in a thermal background. Thus, LPM splitting controls how quickly the particles from Hawking radiation deposit their energy in the first place, while diffusion controls how far this energy spreads. If the diffusion regions from neighboring PBHs overlap and the remaining local equilibration processes are sufficiently rapid, the heating may be approximated as homogeneous. Otherwise, localized hot spots remain around individual PBHs.

\subsection{Implications to Our Results}

We first estimate the time required for the hard cascade. For $\alpha=0.05$, the numerical result from~\cite{He:2024wvt} gives $r_*\simeq3\times10^{11}r_{\mathrm{BH}}$, where $r_*$ is the fitted core radius and $r_{\mathrm{BH}}$ is the black hole radius. Since $r_*\simeq t_{\mathrm{th}}$ in natural units, this corresponds to $t_{\mathrm{th}}\simeq1.5\times10^{-19}$--$1.5\times10^{-17}\,\mathrm{s}$ for $10^8\,\mathrm{g}\leq M\leq10^{10}\,\mathrm{g}$. This interval is many orders of magnitude shorter than the time interval required for a radiation-dominated universe to cool from $30$ to $1\,\mathrm{MeV}$, approximately equals to $0.74\,\mathrm{s}$. The fitted result assumes a massless, gluon-dominated plasma. It therefore shows that the hard cascade is effectively instantaneous within that model, rather than proving complete thermalization in the MeV plasma.

Spatial diffusion occurs on a much longer scale. Using $\lambda_{\mathrm{mfp}}=1/(\alpha^2T)$ with $\alpha=0.05$, fixed $g_*=10.75$, and a radiation-dominated background, we estimate $r_d\simeq 6.22\,\mathrm{cm}$ as the universe cools from $30$ to $1\,\mathrm{MeV}$. This estimate assumes that a thermal perturbation is present throughout the interval and neglects the change in the background caused by PBH evaporation. We next compare this length with the PBH separation using the convention in Sec.\ \ref{sec:massfraction}. If we take $\beta_{30}=1$ and follow the same PBH population to $T=1\,\mathrm{MeV}$, a number-conserving Poisson estimate gives an average nearest-neighbor separation of approximately $0.9$--$4.1\,\mathrm{cm}$ for $10^8\,\mathrm{g}\leq M\leq10^{10}\,\mathrm{g}$. The two lengths are therefore of the same order at $T=1\,\mathrm{MeV}$. However, $\beta_{30}=1$ also changes the expansion history, so this is a comparison of two conditional estimates rather than a self-consistent evolution.

In this estimate, the mean separation between PBHs scales as $\beta_{30}^{-1/3}$. Taking $r_d$ to be comparable to this separation gives $\beta_{30}\simeq 3\times10^{-3}$, $3\times10^{-2}$, and $3\times10^{-1}$ for $M=10^8$, $10^9$, and $10^{10}\,\mathrm{g}$, respectively. Thus, for representative masses below $10^{10}\,\mathrm{g}$, overlap may become important for $\beta_{30}$ of order higher than $10^{-3}$--$10^{-1}$, with a larger value required near the heavier PBH mass. This is a geometric estimate, not a general threshold for homogeneous thermalization or complete volume coverage. For the lowest PBH mass, the distances at $1\,\mathrm{MeV}$ describe the locations of PBHs that have evaporated or are close to evaporation, and the estimate assumes that their deposited heat continues to diffuse. PBH mass loss, entropy injection, the change in the expansion rate, and the particle content of the MeV plasma must be included in a more detailed calculation.

For $M<10^{10}\,\mathrm{g}$, the PBHs evaporate while weak freeze-out and nuclear reactions are active. If neighboring hot regions overlap, our homogeneous calculation can describe the average heating of the plasma. If the hot regions remain separated, the local nuclear reaction rates need not equal the rates evaluated at the average temperature. The precise peaks and valleys in the low-mass abundance curves in Figs.\ \ref{fig04}--\ref{fig07} can therefore depend on diffusion, although the PBH lifetime still determines when the energy is emitted.

For PBHs with $M\gtrsim10^{10}\,\mathrm{g}$, the main effect found in Sec.\ \ref{sec:results} is the increase in the Hubble rate caused by the PBH energy density. PBHs evaporation are overwhelmed by the boost in the Hubble rate. This contribution is present whether or not the particles emitted through Hawking radiation have spread homogeneously. A more detailed treatment of local heating may change the thermal and nuclear corrections, but it does not remove the expansion rate effect that produces the high-mass behavior in Figs.\ \ref{fig04}--\ref{fig07}. Our high-mass results are therefore less sensitive to the details of thermalization. A spatial treatment of the hot regions and their effects on local nuclear reactions remains an important subject for future work.

%------------------------------------------------Summary------------------------------------------------
\subsection{Summary} \label{sec:summary}
We investigated the role of PBHs in BBN based on mean free path arguments. In this picture, photons and neutrinos emitted by PBHs can propagate into the cosmological plasma and participate directly in background nuclear reactions, while heavier particles thermalize within a thick hot envelope surrounding the PBHs, transferring their energy to the plasma and increasing its entropy. As a function of $n_s, \mathrm{d}n_s/\mathrm{d}\ln (k),$ and $M$, we construct dual PBH-BBN models in the \burst code by self-consistently coupling the Hubble expansion rate, entropy density, neutrino heating, and nuclear reaction network together to predict primordial nuclide abundances.

We again stress the key outcome of our analysis: PBH evaporation will substantially increase the comoving entropy density. Previous studies focused on relatively small PBH mass fractions ($\beta\sim10^{-20}$) for which plasma heating from Hawking radiation is negligible, while more recent analyses incorporate the entropy injection explicitly~\cite{Boccia:2024nly}.  In the presence of a larger PBH mass fraction, Fig.\ \ref{fig03} clearly shows the magnitude and epoch of the entropy increase from PBH evaporation. Furthermore, the neutrino component of the Hawking radiation changes the neutron-to-proton interconversion rates as demonstrated in Figs.\ \ref{fig09} -- \ref{fig11}.  Coupled together, the non-standard evolution of the entropy and neutrino densities induce the novel primordial abundance patterns visible in Figs.\ \ref{fig04} -- \ref{fig07}, none of which follow from simple scaling laws over the PBH mass range we study. Our findings highlight the necessity of incorporating entropy evolution self-consistently when modeling the gravitational and nuclear physics of PBH-BBN interactions.

Our results (in Fig.\ \ref{fig04} -- \ref{fig07}) reveal a threshold near $M \approx 10^{10}\,\mathrm{g}$ that separates two distinct regimes of BBN behavior. For $M \leq 10^{10}\,\mathrm{g}$, $\yp$ exhibits oscillatory dependence on PBH mass, while $\xd$ and $\xhe$ are slightly suppressed, and $\xli$ shows modest enhancement. For $M > 10^{10}\,\mathrm{g}$, $\yp$ increases monotonically with $\beta_{30}$, with $\xd$ and $\xhe$ strongly suppressed at large $\beta_{30}$. In this regime, $\xli$ behaves differently, with its maximum variation occurring at an intermediate rather than the highest $\beta_{30}$. Overall, light-element abundances are more sensitive to the PBH fraction $\beta_{30}$ than to the PBH mass $M$ when $M > 10^{10}\,\mathrm{g}$, because the significant increase in the Hubble rate dominates over other effects. For very small PBH mass fractions ($\beta_{30} < 10^{-10}$), the effects on BBN are negligible.

At the threshold around $M \approx 10^{10}\,\mathrm{g}$, we examined how the reaction rates, Hubble expansion rate, and neutron-to-proton ratio collectively determine $\yp$ (from Fig.\ \ref{fig09} to \ref{fig11}). For $M \geq 10^{10}\,\mathrm{g}$, the monotonic increase of $\yp$ with $\beta_{30}$ is primarily driven by the accelerated Hubble expansion caused by the PBH energy density. In contrast, for $M \leq 10^{10}\,\mathrm{g}$, the relationship becomes non-monotonic, influenced by the timing of PBH evaporation and the role of $\beta_{30}$ in modifying reaction rates. The interplay between these effects gives rise to the oscillatory dependence of $\yp$ observed in our simulations.

Large PBH fractions introduce additional challenges to our framework. At $\beta_{30}=1$, the PBH and radiation energy densities are equal at the starting temperature $T=30\,\mathrm{MeV}$. If the PBHs survive as the universe expands, their energy density can subsequently exceed that of radiation. This situation would drive the universe into a matter-dominated phase, conflicting with our assumption of a homogeneous and isotropic background. The associated growth of density inhomogeneities in such a regime cannot be captured within our current treatment. Therefore, while our framework successfully incorporates the key thermal and dynamical effects of PBHs on BBN, it does not account for the nonlinear evolution of large-scale inhomogeneities that may arise in a matter-dominated epoch.

Although our choice of $[n_s, \mathrm{d}n_s/\mathrm{d}\ln (k) ]$ differs from the values inferred from CMB observations, it serves to illustrate how PBHs could influence the processes of BBN under alternative early universe conditions. However, the qualitative effects of PBHs on BBN remain consistent with the mechanisms demonstrated in this work. By introducing an entropy injection mechanism through PBH evaporation, thereby allowing for scenarios that violate entropy conservation, we open a new path to explore how variations in entropy may reshape other aspects of the thermal history of the universe. This framework provides a foundation for future studies examining the interplay between PBH evolution, thermal evolution, and the cosmological conditions that govern the early universe.

%------------------------------------------------References------------------------------------------------
\bibliography{references.bib}% Produces the bibliography via BibTeX.

\end{document}